\documentclass[11pt]{article}
\usepackage{jcapmod}

\usepackage{shorthand}

\NewDocumentCommand{\colornucleus}{omme{_^}}{%
  \begingroup\colorlet{currcolor}{.}%
  \IfValueTF{#1}
   {\textcolor[#1]{#2}}
   {\textcolor{#2}}
    {%
     #3% the nucleus
     \IfValueT{#4}{_{\textcolor{currcolor}{#4}}}% subscript
     \IfValueT{#5}{^{\textcolor{currcolor}{#5}}}% superscript
    }%
  \endgroup
}

\allowdisplaybreaks

\usepackage{array}
\newcolumntype{L}[1]{>{\raggedright\let\newline\\\arraybackslash\hspace{0pt}}m{#1}}
\newcolumntype{C}[1]{>{\centering\let\newline\\\arraybackslash\hspace{0pt}}m{#1}}
\newcolumntype{R}[1]{>{\raggedleft\let\newline\\\arraybackslash\hspace{0pt}}m{#1}}

\usepackage[framemethod=default]{mdframed}
\newmdenv[skipabove=7pt,
skipbelow=7pt,
rightline=false,
leftline=false,
topline=false,
bottomline=false,
backgroundcolor=gray!10,
linecolor=gray,
innerleftmargin=5pt,
innerrightmargin=5pt,
innertopmargin=5pt,
innerbottommargin=5pt,
leftmargin=0cm,
rightmargin=0cm,
linewidth=4pt]{eBox}
\newmdenv[skipabove=7pt,
skipbelow=7pt,
rightline=false,
leftline=false,
topline=false,
bottomline=false,
backgroundcolor=gray!10,
linecolor=gray,
innerleftmargin=5pt,
innerrightmargin=5pt,
innertopmargin=-5pt,
innerbottommargin=5pt,
leftmargin=0cm,
rightmargin=0cm,
linewidth=4pt]{eBox2}
\usepackage[framemethod=default]{mdframed}
\newmdenv[skipabove=7pt,
skipbelow=7pt,
rightline=true,
leftline=true,
topline=true,
bottomline=true,
backgroundcolor=gray!15,
linecolor=gray,
innerleftmargin=5pt,
innerrightmargin=5pt,
innertopmargin=5pt,
innerbottommargin=5pt,
leftmargin=0cm,
rightmargin=0cm,
linewidth=0.75pt]{eBox3}

\usepackage[linktocpage=true]{hyperref}
\hypersetup{
colorlinks=true,
citecolor=darkblue,
linkcolor=reddish,
urlcolor=darkblue,
pdfauthor={},
pdftitle={},
pdfsubject={}
}

\usepackage[normalem]{ulem}

\usepackage{amsmath} % To define example environments
\theoremstyle{definition} % To have non-italic text in examples.

\newcommand{\FF}[4]{\,
{}_2F_1\Bigg[\begin{array}{c}#1,#2\\[2pt] #3 \end{array}\Bigg| \,#4\Bigg]
}

\usepackage{amsthm}
\usepackage{fontawesome5}
\makeatletter
\newlength{\apb@width}
\newcommand{\autoparbox}[2][c]{\settowidth{\apb@width}{#2}\parbox[#1]{\apb@width}{#2}}

\makeatother

\makeatletter
\newlength{\negph@wd}
\DeclareRobustCommand{\negphantom}[1]{%
  \ifmmode
    \mathpalette\negph@math{#1}%
  \else
    \negph@do{#1}%
  \fi
}
\newcommand{\negph@math}[2]{\negph@do{$\m@th#1#2$}}
\newcommand{\negph@do}[1]{%
  \settowidth{\negph@wd}{#1}%
  \hspace*{-\negph@wd}%
}
\makeatother

\newcommand{\github}[1]{%
   \href{#1}{\faGithub}%
}
\newcommand{\reef}[1]{(\ref{#1})}
\allowdisplaybreaks[1]
\usepackage{bbold}
\def\dl{\ud \log}
\begin{document}

\newgeometry{top=2cm, bottom=2cm, left=2cm, right=2cm}

\begin{titlepage}
\setcounter{page}{1} \baselineskip=15.5pt 
\thispagestyle{empty}

\begin{center}
{
\fontsize{20}{18} \bf Cosmological Weight-Shifting Matrices 
}
\end{center}

\vskip 20pt
\begin{center}
\noindent
{\fontsize{14}{18}\selectfont 
Claire de Korte\hs$^{1}$, Harry Goodhew\hs$^{2,3}$, Kamran Salehi Vaziri\hs$^{4}$ and Nicolas Weiss\hs$^{1}$}
\end{center}

\begin{center}
\vskip8pt
\textit{$^1$ Max Planck Institute for Mathematics in the Sciences, Leipzig, 04103, Germany}
  \vskip8pt
\textit{$^2$  Leung Center for Cosmology and Particle Astrophysics,
Taipei 10617, Taiwan}

  \vskip8pt
\textit{$^3$  Center for Theoretical Physics,
National Taiwan University, Taipei 10617, Taiwan}
  \vskip8pt
\textit{$^4$ Institute of Physics, University of Amsterdam, Amsterdam, 1098 XH, The Netherlands}

\end{center}

%=========================================
\vspace{0.8cm}
\begin{center}{\bf Abstract}
\end{center}
\noindent
%Cosmological correlators are challenging to compute.

We construct matrices that shift the scaling dimension of scalar fields for arbitrary de Sitter Feynman diagrams. Acting on a set of master integrals, these weight-shifting matrices 
shift the scaling dimensions of 
individual edges of a given diagram by an integer. They can thus be applied to a broader range of problems and are simpler to implement than earlier derivative-based approaches.
By introducing a Kronecker product representation of 
our matrix formulation, we generalise weight-shifting operators beyond four-point functions to arbitrary tree-level diagrams. 
As an application, we obtain explicit expressions for several massless wavefunction coefficients in four-dimensional de Sitter space, starting from conformally coupled seed functions. Our construction provides a systematic and graph-local approach to generating cosmologically relevant correlators from simpler master integrals.

\end{titlepage}
\restoregeometry

\newpage
\setcounter{tocdepth}{3}
\setcounter{page}{2}

\linespread{1.2}
\tableofcontents
\linespread{1.1}

\newpage

\section{Introduction}\label{sec: introduction}

The accelerated expansion of our universe, both during inflation and in the present cosmological era, is well described by de Sitter space. To describe fluctuations and particle interactions in such a background, one is naturally led to quantum field theory in de Sitter space. A central feature of de Sitter physics is the absence of time-translation invariance. This is directly reflected in the perturbative calculation of cosmological correlators. When computing amplitudes in flat space, energy and momentum conservation trivializes the integrals over interaction vertices. The difficulty of computing these diagrams is then entirely contained in the evaluation of loop integrals. In de Sitter space, however, the absence of energy conservation means that even tree-level diagrams involve non-trivial time integrals. These integrals contain nested products of mode functions, and become especially difficult for fields of arbitrary mass, whose mode functions are generically Hankel functions.

\vskip4pt

Over the past decade, there has been substantial effort to systematize the computation of cosmological correlators \cite{Arkani-Hamed:2015bza,Jazayeri:2022kjy,Benincasa:2022gtd,Arkani-Hamed:2017fdk,Fan:2024iek,Sleight:2019hfp,Benincasa:2019vqr,Pimentel:2022fsc,AguiSalcedo:2023nds,Melville:2021lst,Goodhew:2021oqg}. In particular, the cosmological bootstrap program has provided a way to bypass the direct evaluation of time integrals, using instead basic principles such as symmetry and locality to constrain the form of the answer \cite{Baumann:2020dch,Baumann:2022jpr,Arkani-Hamed:2018kmz,Baumann:2019oyu}. In this approach, cosmological correlators are, instead, seen as solutions to differential equations derived from these principles.
An important result of the bootstrap program is the construction of weight-shifting operators, which relate correlators involving fields with different scaling dimension
and spins~\cite{Costa:2011mg,Costa:2011dw,Costa:2016hju, Simmons-Duffin:2012juh,Karateev:2017jgd,Costa:2018mcg,Iliesiu:2015qra,Baumann:2019oyu,Chen:2023xlt}. These operators are especially useful in cosmology, where massless fields are often the most phenomenologically relevant. In particular, massless correlators can be obtained by acting with weight-shifting operators on correlators of conformally coupled fields, which are much simpler to compute as their mode functions reduce from Hankel functions to exponentials \cite{Arkani-Hamed:2015bza}.

\vskip4pt

More recently, a different, first-order approach, has been developed in analogy with methods used for loop integrals in flat-space scattering amplitudes \cite{Arkani-Hamed:2023kig,Arkani-Hamed:2023bsv,Baumann:2025qjx,Baumann:2024mvm,ampsdiffeqn1,ampsdiffeqn2,ampsdiffeqn3,ampsdiffeqn4,He:2024olr,Glew:2025ypb,Capuano:2025ehm,Westerdijk:2026msm}. This approach organizes the relevant cosmological integrals into a finite-dimensional set of master integrals $\vec I$, which satisfy a first-order differential equation
\begin{equation}
    \ud\vec I = A \cdot \vec I\,.
\end{equation}
Here, $\ud$ is a differential with respect to the kinematic variables, namely the momentum-space variables on which the correlator depends, and $A$ is a matrix of one-forms encoding the singularities of the differential system. The existence of such a finite set of master integrals is closely related to the fact that these integrals can be viewed as generalized Euler integrals~\cite{eulerintegrals}, with connections to twisted cohomology and hyperplane arrangements~\cite{De:2023xue,De:2024zic,McLeod:2026jpz,orlik2013arrangements}.

\vskip4pt

The matrix $A$ governing the differential system for the master integrals can be obtained from simple combinatorial rules acting on graph tubings associated with the underlying Feynman diagram~\cite{Arkani-Hamed:2023kig,Arkani-Hamed:2023bsv,Baumann:2025qjx}. This approach is known as \textit{kinematic flow}.\footnote{We would like to stress that, in spite of the connection with the kinematic flow, no prior knowledge of these rules is needed to understand the present work.} In \cite{massivekinflow}, this framework was generalised from conformally coupled scalar fields to those of arbitrary mass in de Sitter space. Importantly, these combinatorial rules also specify a particular set of master integrals.

\vskip4pt

This naturally raises the question of how weight-shifting applies to these master integrals. In this paper, 
we derive a universal prescription for shifting the scaling dimension associated with both internal and external lines directly at the level of these master integrals. The shift is implemented by acting with a weight-shifting matrix $M$ on the master integrals, mapping the integrals associated with a given weight to the those associated with weight shifted by one
\begin{equation}
    \vec I_{\nu+1}=M\cdot \vec I_\nu\, .
\end{equation}
Here, the weight $\nu$ is related to the scaling dimension of the relevant scalar field by $\Delta=d/2+\nu$.

\vskip4pt

A powerful feature of our construction is that, once the set of master integrals has been defined, weight-shifting becomes a purely algebraic operation. One does not need to take %further
derivatives with respect to the kinematic variables; the shift is implemented by matrix multiplication on the vector of master integrals. Moreover, in contrast to the previous versions of  weight-shifting operators \cite{Arkani-Hamed:2018kmz,Baumann:2019oyu}, which act simultaneously on two different propagators, the matrices constructed here change the weight of individual propagators. In this sense, our construction acts locally on the underlying Feynman graph. This locality is what makes our method straightforward to extend from simple examples to generic diagrams. 
It is also what allows us to shift the weights of multiple propagators by applying weight-shifting matrices iteratively. We can likewise repeatedly shift the weight associated with a single propagator to reach larger and larger weights. When combined with the expansion around conformally coupled mass in~\cite[§5]{massivekinflow}, this allows us to compute wavefunction coefficients for arbitrary (real) weights.
Schematically, for the example of an exchange diagram, weight-shifting matrices can be used to implement the shifts
\vspace{0.42cm}
\ba\nonumber
\adjustbox{valign=c,scale=1}{\includegraphics[page=1]{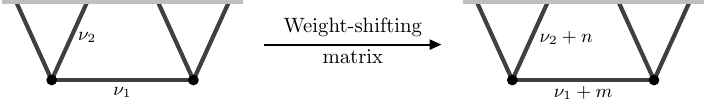}}\, .
    \ea

\vskip5pt

In this paper, we focus on wavefunction coefficients, from which de Sitter late-time correlators can be obtained. However, 
the same construction can also be applied to in-in correlators in de Sitter space, and to momentum-space Witten diagrams in Euclidean Anti-de Sitter (EAdS) space.

\paragraph{Outline}
The paper is organized as follows. In Section \ref{subsec: setting and notation}, we review scalar field theory in de Sitter spacetime, introduce the Feynman rules for wavefunction coefficients, and fix our conventions. In Section \ref{subsec: massive kinematic flow}, we review the construction of the kinematic flow master integrals and the elements of \cite{massivekinflow} that will be used in this paper.
In Section \ref{sec: shifting mass}, we construct weight-shifting matrices in two simple examples; a contact diagram and an exchange diagram. These provide the basic building blocks for the general construction which applies to arbitrary tree-level diagrams as well as loop integrands. To structure this generalisation, in Section \ref{subsec: kronecker notation}, we introduce new, compact notation for the set of master integrals based on Kronecker products. Using this notation, in Section \ref{subsec: generic shifting}, we succinctly construct the generalisation. Finally, in Section \ref{sec: massless}, we apply the resulting weight-shifting matrices in $(3+1)$ spacetime dimensions. In particular, we derive wavefunction coefficients for massless fields from the well-studied wavefunctions of conformally coupled fields.

\vskip4pt
In the appendices, we provide details on explicit computations and place our weight-shifting matrices into mathematical context.
In Appendix~\ref{app:Hyperplane}, we demonstrate how our rules can be understood in the language of twisted cohomology, which connects to the original derivation of the kinematic flow. 
We likewise connect to past work in Appendix~\ref{app:derivativews}, where we review the derivation of the weight-shifting operators that act by taking derivatives and construct operators of this kind that act on our master integrals. In doing this, we demonstrate the agreement between these previously derived differential operators and our matrices.
Next, we provide a small, explicit example, in Appendix~\ref{app: n=2 contact nu shift example}, of the general expressions for the weight-shifting matrices derived in Section~\ref{subsec: generic shifting}. 
We then extend this, deriving an explicit expression for the weight-shifting matrix associated with an arbitrary contact diagram in Appendix~\ref{app: derivation of contact weight-shifting}. This expression does not require one to perform any matrix inversions.
In Appendix~\ref{app:Repeated Shift}, we explore the use of repeated shifts to recover exceptional mass scalars and discuss how this can be used to derive formulae for generic mass scalars using the expansion around conformally coupled masses.
Finally, in Appendix~\ref{app:Massless Alt} we present formulae for the same diagrams considered in~\ref{sec:allmassless} using an alternative, easier to compute, weight-shifting procedure.

\vskip4pt
The weight-shifting matrices derived in this paper can be reproduced using the \textsc{Mathematica} notebook found at \href{https://github.com/CJMDK/Cosmological-Weight-Shifting-Matrices}{\faGithub}.

\section{Massive Fields}\label{sec: master integrals}
In this section, we briefly describe the setting and introduce the notation used throughout this paper. We largely follow the notation of \cite{massivekinflow} and refer the reader to that paper for further details. 
\subsection{Feynman Diagrams in De Sitter}\label{subsec: setting and notation}

Consider scalar fields $\phi_i$ with masses $m_i$ and polynomial interactions, with coupling constants $\lambda_{i_1\cdots i_n}$, in $(d+1)$-dimensional de~Sitter space. Such scalar fields are governed by the action
\begin{equation}\label{eq: action}
S
=-\int \ud^{d+1}x \sqrt{-g}  \left[
\frac12 \,\partial_\mu \phi_i\,\partial^\mu \phi_i
+\frac12\, m_{i}^2\,\phi_i\phi_i
+\sum_{n=3}^{N}\frac{1}{n!}\lambda_{i_1\cdots i_n}\phi_{i_1}\cdots\phi_{i_n}
\right],
\end{equation}
with repeated field indices summed over. Here, $ g_{\mu\nu} $ is the de Sitter metric and  $g$ denotes its determinant. We will be working in the patch of de Sitter space parametrised by planar coordinates, in which the metric takes the form
\begin{equation}
    ds^2=\frac{1}{(H\eta)^2}(-d\eta^2+d\mathbf{x}^2)\,,
\end{equation}
where $\mathbf{x}\in\mathbb{R}^d$ and $\eta\in(-\infty,0)$. 
We are interested in calculating correlation functions of the scalar fields $\phi_i(\eta,\mathbf{x})$ at the late-time boundary $\eta=0$ in the weakly-coupled regime.\footnote{Our method, however, can equally be applied to the calculation of EAdS Witten diagrams in momentum-space.} These correlation functions can be easily constructed from the corresponding wavefunction coefficients \cite{Arkani-Hamed:2018kmz, Baumann:2020dch}, which themselves can be computed using momentum-space Feynman diagrams. The calculation of these Feynman diagrams is the focus of this work.
The Feynman rules we use to compute wavefunction coefficients were first introduced in \cite{massivekinflow}. We briefly review these below.
\vskip4pt
The mode functions associated with the fields $\phi_i$ are solutions to the Euler--Lagrange equation obtained from varying the quadratic part of the action \reef{eq: action} in $\phi_i$. 
We will be working with mode functions that are canonically normalized and satisfy the \textit{Bunch\textendash Davies} initial conditions,
\be
f_\nu(k,\eta)=\frac{\sqrt{\pi H^{d-1}}}{2}\,e^{-i\frac{\pi\nu}{2}}(-\eta)^{d/2}H_{\nu}^{(2)}(-k\eta)\,,
\ee
where $\mathbf{k}$ is a Euclidean vector representing the momentum of the associated particle. We denote the magnitude of the momentum by $k = |\mathbf{k}|$ and refer to it as the ``energy'' of the associated particle.  The parameter $\nu$ is related to the mass of the associated field through\footnote{The scaling dimension is related to mass via $\Delta (d-\Delta)= \frac{m^2}{H^2}$ and consequently to $\nu$ via the relation  $\Delta=\frac{d}{2}+\nu$.}
\begin{equation}
    \frac{m^2}{H^2}=\frac{d^2}{4}+\nu^2\,.
\end{equation}
For the fields to lie in the unitary representations of the de Sitter group $SO(1,d+1)$, we require that $\nu$ either be in the \textit{complementary series} or in the \textit{principal series}. These correspond to the cases $-\frac{d}{2} < \nu < \frac{d}{2}$ and $\nu = i\mu$, with $\mu \in \mathbb{R}$, respectively. Note that $\nu$ is either purely real or purely imaginary. Since the action \reef{eq: action} is invariant under the shadow transformation $\nu \leftrightarrow -\nu$, without loss of generality we can fix the convention $\mathrm{Re}\,\nu \geq 0$ or $\mathrm{Im}\,\nu \geq 0$.
The conformally coupled mass, given by $\frac{m^2}{H^2} = \frac{d^2 - 1}{4}$, corresponds to $\nu = \frac{1}{2}$. For notational convenience, it is useful to define the deviation from this value as $\xi \equiv \nu - \frac{1}{2}$. In this work, we will call $\nu$ the \textit{weight} of the associated field.

\vskip4pt
As described in \cite[§2.2]{massivekinflow}, it is convenient to work in terms of the rescaled mode functions
\ba 
g_\nu(k,\eta) &= - i \sqrt{\frac{\pi}{2}} k^{-\xi} (-k\eta)^\nu H^{(2)}_\nu (-k \eta)\,,\\ \bar{g}_\nu(k,\eta) &= i \sqrt{\frac{\pi}{2}} k^{-\xi} (-k\eta)^\nu H^{(1)}_\nu (-k \eta)\,, 
\ea
and their corresponding Feynman rules. Note that the bar does \emph{not} denote complex conjugation; throughout this work, complex conjugation is denoted by $(\,*\,)$.
For completeness, we summarize the corresponding Feynman rules from \cite[§2.2]{massivekinflow} below: 
\begin{itemize}
    \item \textit{Internal lines:} Each internal line between vertices $i$ and $j$ corresponds to a rescaled bulk-to-bulk propagator
    \begin{equation}
        G_\nu(k,\eta_i,\eta_j)= G_{\nu,\text{F}}(k,\eta_i,\eta_j) - G_{\nu,\text{D}}(k,\eta_i,\eta_j)\,,
    \end{equation}
    where
    \ba
       G_{\nu,\text{F}}(k,\eta_i,\eta_j)&= \bar{g}_\nu(k,\eta_i) g_\nu(k,\eta_j)\theta(\eta_i-\eta_j)+g_\nu(k,\eta_i) \bar{g}_\nu(k,\eta_j)\theta(\eta_j-\eta_i)\,,\\
     G_{\nu,\text{D}}(k,\eta_i,\eta_j)&= g_\nu(k,\eta_i) g_\nu(k,\eta_j)\,, 
    \ea
    are the Feynman and disconnected propagators.
    \item \textit{External lines:} Each external line attached to vertex $i$ corresponds to a rescaled  bulk-to-boundary propagator
    \begin{equation}
        K_\nu(k,\eta_i)= g_\nu(k,\eta_i)\,.
    \end{equation}
    \item \textit{Vertex factor:} 
    To each vertex $i$, we assign a time-dependent coupling\footnote{In this work we will be using $i$ both as an index and to mean $\sqrt{-1}$. Whether it represents an index or $\sqrt{-1}$ will always be clear from context. As a rule, whenever multiplication by $i$ appears, this $i$ is $\sqrt{-1}$.} 
    \begin{equation}
    i \lambda_i H^{-(d+1)}(-\eta_i)^{-(1+\alpha_i+\xi^{(i)})}\,,
    \end{equation}
    and integrate over the interaction time $\eta_i$. Here, $\xi^{(i)}$ denotes the sum of the $\xi$ indices of the lines attached to vertex $i$. We have also defined the parameter
    \be\label{eqn: physical alpha}
    \alpha_i = d - \frac{p_i}{2}(d - 1)\,,
    \ee
    where $p_i$ is the number of lines connecting to vertex $i$. For fixed $d$, note that $\alpha_i$ just quantifies a geometric characteristic of the underlying Feynman diagram and 
    does not depend on the masses of the fields.
    Moreover, whilst, in principle, $\alpha_i$ is determined by the diagram in question, it is often convenient to treat it as an independent complex variable. We only set $\alpha_i$ to be the value prescribed by the diagram at the end of a calculation.

\end{itemize}
Below we illustrate the use of these Feynman rules in the case of a contact and exchange diagram.

%They will become our main examples to demonstrate weight shifting.

\vskip8pt

\noindent\textbf{Contact diagram}\newline
    \noindent For a contact diagram with $p$ external lines, we have
    \ba
\adjustbox{valign=c,scale=1}{\includegraphics[page=1]{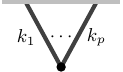}} 
        = \frac{i\lambda}{H^{d+1}}\int \frac{\ud\eta} {(-\eta)^{1+\alpha + \xi_1+\cdots+\xi_p}}\prod_{i=1}^pK_{\nu_i}(k_i,\eta)\,,
    \ea
    where $\lambda$ is the coupling for such an interaction and $k_i$ and $\nu_i$ are the energy and weight corresponding to $i$-th line.

    \vskip8pt

\noindent\textbf{Exchange diagram}\newline
    \noindent For the four-point exchange diagram, with one internal line and two vertices, we have
    \ba\label{eq: exchange}
\adjustbox{valign=c,scale=1}{\includegraphics[page=2]{img/TikzPictures.pdf}}&= \int \frac{i\lambda_1H^{-d-1}\ud\eta_1 }{(-\eta_1)^{1+\alpha_1+\xi_1+\xi_2+\xi_{12}}}
\frac{i\lambda_2H^{-d-1}\ud\eta_2 }{(-\eta_2)^{1+\alpha_2+\xi_3+\xi_4+\xi_{12}}}
 G_{\nu_{12}}(Y,\eta_1,\eta_2)\\
 &\times K_{\nu_1}(k_1, \eta_1)K_{\nu_2}(k_2, \eta_1)K_{\nu_3}(k_3, \eta_2)K_{\nu_4}(k_4, \eta_2)\,.
\ea
The energy and weights of the external lines are labelled by $k_i$  and $\nu_i$, respectively, while the internal (exchange) line is labelled by $Y$ and $\nu_{12}$. Since the number of edges at each vertex is $p_1=p_2=3$, we have $\alpha_1=\alpha_2=\frac{3-d}{2}$. Finally, $\lambda_1$ and $\lambda_2$ are the couplings corresponding to the interactions at vertices $1$ and $2$.

\vskip8pt

Before moving on, we make a few brief remarks about these Feynman rules and some of the terminology used in this work. Formally, wavefunction coefficients, like amplitudes, are defined by an infinite sum over the integrals defined by Feynman diagrams. For example, it is often necessary to sum over the contributions from various channels. However, we will also refer to the contribution to the wavefunction coefficient coming from a single Feynman diagram as simply the wavefunction coefficient, it should be clear which definition of the wavefunction coefficient is being used from context.

\vskip4pt

Since the Feynman rules presented above are given in terms of the rescaled mode functions $g_\nu$ and $\bar{g}_\nu$, they calculate rescaled versions of the standard wavefunction coefficients considered in cosmology. The factor introduced by this rescaling is made explicit in \cite[equation 2.19]{massivekinflow}. Furthermore, as discussed in \cite[§4]{massivekinflow}, the contributions to wavefunction coefficients arising from the disconnected propagator $G_{\nu,D}$ are themselves products of lower-point diagrams. The factors appearing in these disconnected contributions correspond to diagrams of reduced complexity when compared with the original diagram. For example, in the case of the exchange diagram \eqref{eq: exchange}, the disconnected piece can be written as a product of two contact diagrams. Consequently, in what follows we restrict our attention to the fully connected part of diagrams (the part with no $G_{\nu,D}$ contributions). To avoid writing `fully connected component' repeatedly, henceforth the term Feynman diagram will be used to mean `fully connected component of the Feynman diagram'. Finally, to avoid clutter we keep the factors of $(i\lambda_i)$ prescribed by the Feynman rules implicit. 
These can easily be recovered by taking all the vertices of the relevant Feynman diagram into account.

\subsection{Master Integrals}\label{subsec: massive kinematic flow}
\noindent Instead of evaluating the integrals explicitly, one can alternatively compute the function associated with a given Feynman diagram by considering the system of differential equations it satisfies. After imposing appropriate boundary conditions, solving these differential equations gives an expression for the function, without needing to do any integration. This method has been used extensively for computing the Feynman integrals that appear in scattering amplitudes \cite{ampsdiffeqn1,ampsdiffeqn2,ampsdiffeqn3,ampsdiffeqn4}. Recently, this approach was applied to arbitrary mass de Sitter wavefunction coefficients in \cite{massivekinflow}, where these differential equations are derived from simple graphical rules called \textit{kinematic flow}. The procedure outlined in that paper introduces the four auxiliary functions
\ba\label{eq: h def}
    h^\pm_\nu(k,\eta)&=\frac{1}{2}\left[g_\nu(k,\eta)\mp\frac{i}{k}\partial_\eta g_\nu(k,\eta)\right],
     \\
     \bar{h}^\pm_\nu(k,\eta)&=\frac{1}{2}\left[\bar{g}_\nu(k,\eta)\mp\frac{i}{k}\partial_\eta \bar{g}_\nu(k,\eta)\right].
\ea
From these functions, we define a set of \textit{master integrals} by replacing $g_\nu$ and $\bar{g}_\nu$ in our Feynman rules with $h_\nu^\pm$ and $\bar{h}_\nu^\pm$, respectively. 
We label the master integrals associated with a particular Feynman diagram as $\psi^{\pm\cdots\pm}$, the $(\,\pm\,)$ specifying the sequence of signs on the  $h_\nu^\pm$ and $\bar{h}^\pm_\nu$ appearing in the integrand. To ensure that our set of master integrals is closed under differentiation with respect to the kinematic variables, we include \emph{collapsed} functions in which we have replaced bulk-to-bulk propagators with $\delta$-functions. We label these $J$.  Note that, given a set of master integrals, the corresponding
wavefunction coefficient is just the sum of the non-collapsed integrals. We refer the reader to \cite[§3.1]{massivekinflow} for further details. We give an explicit formula  for the set of master integrals associated with an arbitrary tree-level Feynman diagram in Section~\ref{subsec: kronecker notation}.

\vskip4pt

The set of master integrals associated with a given Feynman diagram forms a closed system under differentiation with respect to kinematic variables. By arranging the master integrals as entries of a vector $\vec{I}$, they are shown in \cite{massivekinflow} to satisfy a differential equation of the form
\begin{equation} \label{eq: Kinematic Flow}
    \mathrm{d}\vec{I}=A\cdot\vec{I}\,,
\end{equation}
where $(\,\cdot\,)$ denotes matrix multiplication. In what follows, we will refer to this matrix $A$ as \emph{the} $A$-matrix of a given Feynman diagram. These $A$-matrices can be constructed using a simple set of rules introduced in \cite{massivekinflow}. Our main results do not rely on the explicit forms of these $A$-matrices, just on some of the properties they have. Consequently, we will not review the rules for the construction of $A$-matrices here, instead referring the reader to \cite[§4.2]{massivekinflow} for these.

\vskip4pt
 
The fact that the integrals in $\vec{I}$ form a closed system under differentiation is a direct consequence of the following differential equations
\begin{align}
    \label{eqn: h differential equations}
    \partial_\eta h_\nu^{\pm}(k,\eta)
    &= \pm i k \, h_\nu^{\pm}(k,\eta)
    \pm \frac{\xi}{\eta}
    \left[ h_\nu^{+}(k,\eta) - h_\nu^{-}(k,\eta) \right], \\
    \partial_k h_\nu^{\pm}(k,\eta)
    &= \pm i \eta \, h_\nu^{\pm}(k,\eta)
    - \frac{\xi}{k} \, h_\nu^{\mp}(k,\eta)\,, \label{eqn: dk of h differential equations}
\end{align}
where an identical set of equations holds for $\bar{h}^\pm_\nu$. 
In the present work, our construction of weight-shifting matrices is achieved using the following crucial identity
\ba\label{eqn: h shift relations}
    h^\pm_{\nu+1}(k,\eta) = \mp i\eta \,h^\pm_\nu(k,\eta) + \frac{\nu}{k}[h_\nu^+(k,\eta)+h_\nu^-(k,\eta)]\,.
\ea
As with the differential equations, identical shift relations hold for the $\bar{h}^\pm_\nu$. These  identities follow directly from shift relations satisfied by Hankel functions~\cite[§10.6]{DLMF},
\ba\label{eq: hankel shift relation}
    \partial_z H_{\nu}(z) =\pm \frac{\nu}{z}  H_{\nu} (z) \,\mp\,  H_{\nu\pm 1}(z)\,.
\ea
The superscript denoting the kind of Hankel function has been dropped as both kinds satisfy this equation. These elementary shift relations are the building blocks of all our weight-shifting operators.

%%%%%%%%%%%%%%%%%%%%%%%%%%%%%%%%%%%%%%%%%%%%%%%%%%%%%%%%%%%%%%%%%
\subsubsection*{Conformally coupled limit}
For completeness, we briefly discuss the case of conformally coupled fields, for which much of the formalism introduced above simplifies dramatically. As mentioned in the introduction, conformally coupled fields have weight  $\nu=\half$ or equivalently $\xi=0$. For this value of the weight, the $h$-functions are 
\begin{equation}
    h^-_{1/2}(k,\eta)=\bar{h}^+_{1/2}(k,\eta)=0\,, \quad h^+_{1/2}(k,\eta)=(\bar{h}^-_{1/2}(k,\eta))^*=e^{ik\eta}\,,
\end{equation}
This leads to many simplifications. First of all, since $h_{1/2}^-$ and $\bar{h}_{1/2}^+$ vanish, the number of non-zero master integrals decreases substantially. Moreover, the contributions to master integrals arising from conformally coupled bulk-to-boundary propagators can be collected into a single exponential. Explicitly, for $n$ conformally coupled bulk-to-boundary propagators, each attached to the same vertex and each labelled by an energy $k_i$, we have
\begin{equation}\label{eq: conformal limit total energy}
    \prod_{i=1}^n h^+_{1/2}(k_i,\eta) =\prod_{i=1}^n e^{i \eta  k_i}= {e^{i \eta X}}\,, \qquad \text{where}\quad X\equiv \displaystyle{\sum_{i=1}^n k_i}\,.
\end{equation}
From now on, we use $X_i$ to denote the total energy of the conformally coupled bulk-to-boundary propagators attached to vertex $i$.

\vskip4pt

For example, recall the exchange diagram \eqref{eq: exchange} from Section~\ref{subsec: setting and notation}, and set the weights of all the external lines to be conformally coupled. The corresponding master integrals are given, in terms of $X_1=k_1+k_2$ and $X_2=k_3+k_4$, by
\ba\label{eqn: exchange conf coup legs}
\adjustbox{valign=c,scale=1}{\includegraphics[page=3]{img/TikzPictures.pdf}}:
\psi^{\pmr\pmb}= \int \frac{\ud\eta_1 }{(-\eta_1)^{1+\alpha_1+\xi}}
\frac{d\eta_2 }{(-\eta_2)^{1+\alpha_2+\xi}}
 e^{i(X_1 \eta_1 + X_2\eta_2)}G^{\pmr\pmb}_\nu(Y,\eta_1,\eta_2)\,,
\ea
where we have defined a new, modified, bulk-to-bulk propagator which arises from the replacement of $g_\nu$ and $\bar{g}_\nu$ with $h^\pm_\nu$ and $\bar{h}^\pm_\nu$ in the Feynman propagator. Explicitly, we have
\begin{equation}\label{eqn: G definition}
    G_\nu^{\textcolor{red}{\pm}\textcolor{blue}{\pm}}(Y,\eta_1,\eta_2)=\bar{h}_\nu^{\textcolor{red}{\pm}}(Y,\eta_1)h_\nu^{\textcolor{blue}{\pm}}(Y,\eta_2)\theta(\eta_1-\eta_2)+h_\nu^{\textcolor{red}{\pm}}(Y, \eta_1)\bar{h}_\nu^{\textcolor{blue}{\pm}}(Y,\eta_2)\theta(\eta_2-\eta_1)\,,
\end{equation}
with exchange energy given by $Y=|\mathbf{k}_1+\mathbf{k}_2|=|\mathbf{k}_3+\mathbf{k}_4|$. 

\vskip4pt

From now on, we use the colour light grey to represent external lines with conformally coupled weights.

%%%%%%%%%%%%%%%%%%%%%%%%%%%%%%%%%%%%%%%%%%%%%%%%%%%%%%%%%%%%%%%%%

%%%%%%%%%%%%%%%%%%%%%%%%%%%%%%%%%%%%%%%%%%%%%%%%%%%%%%%%%%%%%%%%%
\subsection{Shifting Vertex Parameter}\label{subsubsec: A-matrix alpha shift}
We now discuss the main result from \cite{massivekinflow} that we will use in this paper. Recall the $A$-matrix introduced in \reef{eq: Kinematic Flow}. Given a Feynman diagram, its $A$-matrix encodes a set of linear relations between master integrals and their derivatives with respect to the kinematic variables. The entries of the $A$-matrix are $\ud \log$-forms in the kinematic variables of the diagram. We denote the $\partial_{k_i}$-component of the $A$-matrix by
\be
A^{(k_i)} \equiv A\Big|_{\partial k_i}\,.
\ee
Equivalently, one can obtain $A^{(k_i)}$ by setting $dk_i=1$ and $dk_{j}=0$ for $j\neq i$ in the $A$-matrix. 

\vskip4pt

Now suppose the Feynman diagram has a vertex $i$ and corresponding vertex parameter $\alpha_i$. For simplicity, assume that the vertex $i$ has some non-zero number of conformally coupled bulk-to-boundary lines attached to it. By \eqref{eq: conformal limit total energy}, the diagram's master integrals are then functions of the total energy associated with these lines, $X_i$. Later, we will return to the case where there are no conformally coupled external lines.  Use $\vec{I}_{\alpha_i}$ to denote the vector of master integrals corresponding to this Feynman diagram, and $A_{\alpha_i}^{(X_i)}$ to denote the $\partial_{X_i}$-component of its $A$-matrix. 
In the case where $\xi=0$, an immediate consequence of the differential equation \eqref{eqn: dk of h differential equations} is that one can use the $A$-matrix to shift the value of the vertex parameter $\alpha_i$, 
\begin{equation}\label{eqn: introducing alpha shift}
    -i\vec{I}_{\alpha_i}=A^{(X_i)}_{\alpha_i+1}\cdot\vec{I}_{\alpha_i+1}\,,
\end{equation}
where $\vec{I}_{\alpha_i+1}$, as the notation suggests, is just $\vec{I}_{\alpha_i}$, where $\alpha_i$ has been replaced by $\alpha_i+1$ while all the other vertex parameters are kept the same. Likewise for $A_{\alpha_i+1}^{(X_i)}$. See the derivations in \cite[§3]{massivekinflow} for explicit calculations that demonstrate this shifting property. For clarity, we emphasize that the dependence on the weights $\nu$ is the same in both $A^{(X_i)}_{\alpha_i}$ and $A^{(X_i)}_{\alpha_i+1}$. To avoid introducing cumbersome notation, we keep the dependence of these matrices on the weights implicit where there is no ambiguity. Schematically, the shift in vertex parameter induced by the action of an $A$-matrix on an exchange diagram can be represented as follows
\vspace{0.2cm}
\ba\nonumber
\adjustbox{valign=c,scale=1}{\includegraphics[page=1]{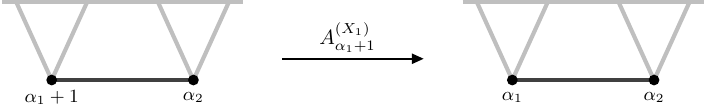}} \,.
    \ea

\vskip4pt

One may similarly attempt to derive a shift relation for $\alpha_i$ analogous to \reef{eqn: introducing alpha shift}, using the component of $A$ associated with the energy corresponding to a different line attached to vertex $i$. In this case, however, one must take the mixing between master integrals induced by \eqref{eqn: dk of h differential equations} into account. This makes the derivation of any potential shift substantially more involved. Naively, this suggests that it may be difficult to obtain such a relation for diagrams in which no conformally coupled external line is attached to vertex $i$. This difficulty can be bypassed with the following trick. Attach a virtual\footnote{Here, virtual means that adding this line does not change the value of the $\alpha_i$ parameter.} external line with energy $X_i$ to vertex $i$ and then perform the shift of $\alpha_i$ using $A^{(X_i)}_{\alpha_i+1}$. Since the contribution of this virtual line to the integrand is given by $h^+_{1/2}(X_i,\eta_i)=e^{i X_i \eta_i}$, setting $X_i\to 0$ translates to setting $e^{i X_i \eta_i}\to 1$, and one recovers the desired shifting relation.

\section{Shifting Mass}
\label{sec: shifting mass}

Having described the setting and relevant preliminaries, we now present our main results. We start by carrying out step-by-step derivations of weight-shifting matrices on two simple examples, a contact diagram and a single-exchange diagram. Kronecker products are then used to introduce new notation for the set of master integrals of a given tree-level Feynman diagram. We then demonstrate the power of this notation in allowing us to succinctly ``uplift'' the calculations performed on the simple examples to arbitrary tree-level diagrams.

\subsection{Contact Diagram}\label{subsec: contact diagram}

We begin by considering possibly the simplest example, namely a %three-point function
Feynman diagram with two conformally coupled lines and one arbitrary mass line. This is given by the integral
\beq
 \psi_{\alpha,\nu}= \adjustbox{valign=c,scale=1}{\includegraphics[page=4]{img/TikzPictures.pdf}}
=  \int \frac{\ud\eta}{ (-\eta)^{1+\alpha + \dev}}\, e^{iX\eta} \,  g_\nu (k,\eta)\,,
\label{eq: psi def contact}
\eeq
where $k\equiv k_1$,  $X\equiv k_2+k_3$ and $\nu =\half+\xi$ and we have made the $\alpha$ and $\xi$ dependence explicit to conform with notation introduced later for more general diagrams. By replacing $g_\nu$ with $h^+_\nu$ and $h^-_\nu$, we construct the master integrals of this diagram
\be\label{eq: contact masters}
\psi_{\alpha,\nu}^\pm=\int \frac{\ud \eta}{(-\eta)^{1+\alpha+\xi}} e^{i X\eta } h^{\pm}_\nu(k,\eta)\,.
\ee
From the definition of the functions $h_\nu^\pm(k,\eta)$ in \reef{eq: h def}, it is clear that $\psi_{\alpha,\nu}=\psi_{\alpha,\nu}^++\psi_{\alpha,\nu}^-$.
It is helpful to arrange these master integrals as a vector
\be
\vec I_{\alpha,\nu} =  \begin{bmatrix} \psi_{\alpha,\nu}^+ \\\psi_{\alpha,\nu}^-
\end{bmatrix}.
\ee
Below we derive two matrices, both of which shift the weight of each of the master integrals, $\nu$, by one. The first, $M$, does so whilst leaving $\alpha$ constant. The second, $\mathcal{M}$, also shifts $\alpha$ by minus one. More precisely, we construct two $2\times2$ matrices $M$ and $\mathcal{M}$ satisfying
\be\label{eq:M def}
\vec I_{\alpha,\nu+1}= {M} \cdot \vec I_{\alpha,\nu}\quad\text{ and }\quad\vec I_{\alpha,\nu+1}= \mathcal{M} \cdot \vec I_{\alpha+1,\nu}\,,
\ee
where
\be\label{eq: M for contact}
i M=  \begin{bmatrix}
        X^+ & X^- \\
        X^+& X^- 
    \end{bmatrix} -
    \begin{bmatrix}
        1 & 0 \\
        0 & -1
    \end{bmatrix}, \quad \mathcal{M}=\begin{bmatrix}
    \frac{\nu}{k}-\frac{\alpha+1}{X+k} &&\frac{\nu}{k}-\frac{\xi}{X+k}\\
    \frac{\nu}{k}+\frac{\xi}{X-k}&&\frac{\nu}{k}+\frac{\alpha+1}{X-k}
\end{bmatrix},
\ee
in which we have defined $X^{\pm}=\frac{\nu}{k}\frac{X\pm k}{1+\alpha+\xi}$. The reasoning behind the introduction of $\mathcal{M}$ is made clear in the derivation below.

\vskip4pt

\noindent Using the matrix $M$, the wavefunction coefficient $\psi_{\alpha,\nu+1}$ is then given by the sum of the elements of the shifted vector $\vec{I}_{\alpha,\nu+1}$,
\ba\label{eq: contact full}
    \psi_{\alpha,\nu+1}=\frac{i}{k(1+\alpha+\xi)}\left[k(\alpha-\xi)(\psi^+_{\alpha,\nu}-\psi^-_{\alpha,\nu})-2\nu X (\psi^+_{\alpha,\nu}+\psi^-_{\alpha,\nu})\right].
\ea
Similarly, using the matrix $\mathcal{M}$, we find
\be\label{eq: contact weightshifted}
\psi_{\alpha,\nu+1}=\left(\frac{2\nu}{k}-\frac{\alpha+1}{X+k}+\frac{\xi}{X-k}\right)\psi_{\alpha+1,\nu}^{+}+\left(\frac{2\nu}{k}-\frac{\xi}{X+k}+\frac{\alpha+1}{X-k}\right)\psi_{\alpha+1,\nu}^{-}\,.
\ee
In the following inset, we present a derivation of the weight-shifting matrices. For those more accustomed to the twisted cohomology approach in the language of hyperplane arrangements we discuss the equivalent derivation in this language in Appendix~\ref{app:Hyperplane}.

\vspace{0.1cm}
\begin{eBox3}
{\bf Derivation:}\\
\textit{Shifting $\nu$}: First, using the identity \reef{eqn: h shift relations}, we can relate the master integrals with weights $\nu$ and $\nu+1$,
    \ba\psi_{\alpha,\nu+1}^\pm&=\int \frac{\ud \eta}{(-\eta)^{2+\alpha+\xi}} e^{i X\eta } h^{\pm}_{\nu+1}(k,\eta)\\
     &= \pm i\, \psi_{\alpha,\nu}^\pm + \frac{\nu}{k}\left(\psi_{\alpha+1,\nu}^+ +\psi_{\alpha+1,\nu}^- \right).
    \ea
    As a matrix equation, this is
    \be\label{eq: intermediate I}
    \vec I_{\alpha,\nu+1} = i B \cdot \vec I_{\alpha,\nu} +\frac{\nu}{k} C \cdot \vec I_{\alpha+1,\nu}\,,
    \ee
    where we have defined the matrices
    \be\label{eq: BC def}
    B=\begin{bmatrix}1 & 0\\
    0& -1 \end{bmatrix},\qquad 
    C= \begin{bmatrix}1 & 1\\
    1& 1 \end{bmatrix}.
    \ee
    Three distinct vectors appear in~\reef{eq: intermediate I}. We want to derive relationships between just two vectors. Recall from~\eqref{eqn: introducing alpha shift} that the $\partial_X$-component of the $A$-matrix associated with this diagram, $A^{(X)}$, can be used to shift $\alpha$. With this in mind, there are two ways to proceed.
    \begin{enumerate}
        \item \emph{Equal $\alpha$}: We use the inverse of $A^{(X)}$ to replace $\vec{I}_{\alpha+1,\nu}$ with $\vec{I}_{\alpha,\nu}$,
        \be\label{eq: xi+1 with A-1}
        \vec I_{\alpha,\nu+1} = i \left( B - \frac{\nu}{k} \, C \cdot \left[A^{(X)}_{\alpha+1}\right]^{-1}  \right)\cdot \vec I_{\alpha,\nu}.
        \ee
        \item \emph{Shifted $\alpha$}: Rather than inverting $A^{(X)}$, we use it as it is to replace $\vec{I}_{\alpha,\nu}$ with  $\vec{I}_{\alpha+1,\nu}$,
        \be
        \label{eq: alpha nu shift matrix}
        \vec I_{\alpha,\nu+1} =  \left( - B \cdot A^{(X)}_{\alpha+1}+ \frac{\nu}{k} \, C \right)\cdot \vec I_{\alpha+1,\nu}\,.
        \ee
        In practice, it is often easier to perform this kind of weight-shifting than weight-shifting in which $\alpha$ is left unchanged. This is because \cite{massivekinflow} provides a simple algorithm for constructing $A$-matrices, but inverting $A$-matrix components can be non-trivial. 
    \end{enumerate}
Using the algorithm defined in \cite{massivekinflow}, the $A$-matrix for this diagram and its $\partial_X$-component with $\alpha$ replaced by $\alpha+1$ are given by
     \be
     A_\alpha=  \begin{bmatrix}
         \alpha\,\dl(X+k) & \xi \, \dl\left(\frac{X+k}{k}\right) \\
         \xi \,\dl\left(\frac{X-k}{k}\right) & \alpha\,\dl(X-k) 
     \end{bmatrix},\quad\; A^{(X)}_{\alpha+1}= \begin{bmatrix}
         \frac{\alpha+1}{X+k}& \frac{\xi}{X+k}  \\
         \frac{\xi}{X-k} & \frac{\alpha+1}{X-k}
     \end{bmatrix}.
     \ee
     Substituting the matrix on the right and its inverse into \reef{eq: alpha nu shift matrix} and \reef{eq: xi+1 with A-1}, we arrive at the expressions \reef{eq: M for contact} for the weight-shifting matrices. 
\end{eBox3}

\vspace{0.1cm}

Before moving to the second example, we illustrate explicitly how this weight-shifting matrix works. Directly performing integral \reef{eq: psi def contact} gives the following expression for the diagram:
\be\label{eq:Explicit Contact}
\psi_{\alpha,\nu}= C_{\alpha,\nu}\left(1-\frac{\xi}{\alpha}\right)\,{}_2F_1\bigg[\begin{array}{c}
-\alpha-\dev\ ,\  1-\alpha+\dev\\[-1pt]
1-\alpha
\end{array}\bigg\rvert \, \frac{k-X}{2k}\,\bigg]\,,
\ee
where
\be\label{eq:Contact Coeff}
C_{\alpha,\nu}=(2k)^\alpha e^{\frac{i\pi}{2}(\alpha+\xi)} \frac{\Gamma(-\alpha+\xi)\Gamma(-\alpha-\xi)}{\Gamma(-\alpha)}\,.
\ee
Similarly, one can directly evaluate the master integrals in \reef{eq: contact masters} to find 
\ba
\psi^+_{\alpha,\nu} &= C_{\alpha,\nu}  \; {}_2F_1\bigg[\begin{array}{c}
-\alpha-\dev\ ,\  -\alpha+\dev\\[-1pt]
-\alpha
\end{array}\bigg\rvert \, \frac{k-X}{2k}\,\bigg] \,,\\
\psi^-_{\alpha,\nu} &= - \frac{\xi}{\alpha}C_{\alpha,\nu} \;  {}_2F_1\bigg[\begin{array}{c}
-\alpha-\dev\ ,\  -\alpha+\dev\\[-1pt]
1-\alpha
\end{array}\bigg\rvert \, \frac{k-X}{2k}\,\bigg] \,.
\ea
 Using these relations, the weight-shifting relation provided in \reef{eq: contact full} follows from the hypergeometric identity
 \ba
 \FF{a-1}{b+2}{\frac{a+b}{2}+1}{z}
&=-\frac{(a+b)\left[(a-b-1)(1-2z)-b\right]}{2b(b+1)}
\FF{a}{b}{\frac{a+b}{2}}{z}\\
&\quad+
\frac{(a-b)\left[(a-b-1)(1-2z)+b\right]}{2b(b+1)}
\FF{a}{b}{\frac{a+b}{2}+1}{z}\,,
\ea
which is derived from Gauss's contiguous relations~\cite{vidunas2003contiguous}. In this sense, our weight-shifting matrices play a role analogous to the contiguous relations known for generalized hypergeometric functions.

%%%%%%%%%%%%%%%%%%%%%%%%
\subsection{Exchange Diagram}\label{subsec: exchange example}

Above we demonstrated how the identity~\reef{eqn: h shift relations} and the relevant $A$-matrix can be used to derive weight-shifting matrices in the case of a simple contact diagram. We now extend this analysis to the diagram depicted in \eqref{eqn: exchange conf coup legs}, an exchange diagram with conformally coupled external lines and a single arbitrary mass internal line.

\vskip4pt

As described above, the master integrals corresponding to this diagram are given by
\be
\psi_{\alpha_1,\alpha_2,\nu}^{\pmr\pmb} =
  \int \ud\eta_1\ud\eta_2 \,\frac{e^{iX_1\eta_1}e^{iX_2\eta_2} }{(-\eta_1)^{1+\alpha_1+\dev}(-\eta_2)^{1+\alpha_2+\dev}}  G^{\pmr\pmb}_\nu(Y,\eta_1,\eta_2)\,,
\ee
where $\nu=\half+\xi$ is the weight of the internal line and $\alpha_1$, $\alpha_2$ are vertex parameters of the first and second vertices. The wavefunction coefficient associated with this diagram is then given by the sum
\begin{equation}
    \psi_{\alpha_1,\alpha_2,\nu}= \psi_{\alpha_1,\alpha_2,\nu}^{++}+\psi_{\alpha_1,\alpha_2,\nu}^{-+}+\psi_{\alpha_1,\alpha_2,\nu}^{+-}+\psi_{\alpha_1,\alpha_2,\nu}^{--}\,.
\end{equation}
For the set of master integrals to be closed under differentiation, we must introduce one further integral. This integral, called a source function, corresponds to a collapse in the bulk-to-bulk propagator and is given by \cite{massivekinflow}
\be
 J_{\alpha_1,\alpha_2}=  \int \frac{\ud\eta}{(-\eta)^{1+\alpha_1+\alpha_2}} \,e^{i(X_1+X_2)\eta} \,.
\ee
We can now organise the set of master integrals associated with the diagram as a vector
\be\label{eq: shift exchange}
\vec I_{\alpha_1,\alpha_2,\nu} =
 [\psi^{++}_{\alpha_1,\alpha_2,\nu} \ \ \
   \psi^{+-}_{\alpha_1,\alpha_2,\nu} \ \ \
  \psi^{-+}_{\alpha_1,\alpha_2,\nu} \ \ \
  \psi^{--}_{\alpha_1,\alpha_2,\nu} \ \ \
 J_{\alpha_1,\alpha_2}]^T\,.
\ee
As in the previous section, below we derive two matrices $M$ and $\mathcal{M}$ which realise the shifts
\be
\quad \vec I_{\alpha_1,\alpha_2,\nu+1} = M \cdot \vec I_{\alpha_1,\alpha_2,\nu}\quad\text{ and }\quad\vec I_{\alpha_1,\alpha_2,\nu+1} = \mathcal{M} \cdot \vec I_{\alpha_1+1,\alpha_2+1,\nu}{\,.}
\ee
Explicitly, these matrices are given by
\ba\label{eq: M for exchange}
M= -{\small\begin{bmatrix}
        (X_1^+-1)(X_2^+-1) & X_2^-(X_1^+-1) & X_1^-(X_2^+-1) & X_1^-X_2^- & L_0 \\
         X^+_2(X_1^+-1) &(X_1^+-1)(X_2^-+1)& X_1^-X_2^+&X_1^-(X_2^-+1)&L_0 \\
        X_1^+(X_2^+-1) & X_1^+X_2^- & (X_1^-+1)(X_2^+-1) & X_2^-(X_1^-+1) & L_0 \\
         X_1^+X_2^+&X_1^+(X_2^-+1)& X_2^+(X_1^-+1)&(X_1^-+1)(X_2^-+1)& L_0\\
         0& 0& 0& 0& -1
    \end{bmatrix}
    },
\ea
and
\begin{equation}\label{eqn: curly M for exchange}
\mathcal{M}={\small\begin{bmatrix}Q_1^+Q_2^+&Q_1^+P_2^+&P_1^+Q_2^+&P_1^+P_2^+&R^{++}\vphantom{\frac{-\xi}{(X_1+Y)(X_2+Y)}}\\Q_1^+P_2^-&Q_1^+Q_2^-&P_1^+P_2^-&P_1^+Q_2^-&R^{+-}\vphantom{\frac{-\xi}{(X_1+Y)(X_2+Y)}}\\P_1^-Q_2^+&P_1^-P_2^+&Q_1^-Q_2^+&Q_1^-P_2^+&R^{-+}\vphantom{\frac{-\xi}{(X_1+Y)(X_2+Y)}}\\P_1^-P_2^-&P_1^-Q_2^-&Q_1^-P_2^-&Q_1^-Q_2^-&R^{--}\vphantom{\frac{-\xi}{(X_1+Y)(X_2+Y)}}\\0&0&0&0&0\vphantom{\frac{-\xi}{(X_1+Y)(X_2+Y)}}\end{bmatrix}+\begin{bmatrix}0&0&0&0&\frac{-\xi}{(X_1+Y)(X_2+Y)}\\0&0&0&0&\frac{X_2(1+\alpha_2)-X_1(1+\alpha_1)-Y(2+\alpha_1+\alpha_2)}{(X_1+X_2)(X_1+Y)(X_2-Y)}\\
0&0&0&0&\frac{X_1(1+\alpha_1)-X_2(1+\alpha_2)-Y(2+\alpha_1+\alpha_2)}{(X_1+X_2)(X_1-Y)(X_2+Y)}\\0&0&0&0&\frac{\xi}{(X_1-Y)(X_2-Y)}\\
0&0&0&0&-\frac{(1+\alpha_1+\alpha_2)(2+\alpha_1+\alpha_2)}{(X_1+X_2)^2}
\end{bmatrix}},
\end{equation}
where we have defined
\begin{alignat}{2}
    X^{\pm}_i&= \frac{\nu}{Y}\frac{X_i\pm Y}{1+\alpha_i+\xi}\,, \quad L_0 &&= \frac{\nu}{Y}\frac{X_1+X_2}{(1+\alpha_1+\xi)(1+\alpha_2+\xi)}\,,\\
    Q^\pm_i&=\frac{\nu}{Y}\mp\frac{1+\alpha_i}{X_i\pm Y}\,, \quad P_i^\pm&&=\frac{1}{2Y}+\frac{\xi X_i}{Y(X_i\pm Y)}\,, \quad R^{\textcolor{red}{\pm}\textcolor{blue}{\pm}}=\frac{\nu}{Y}\left[\frac{1}{X_1\textcolor{red}{\pm}Y}+\frac{1}{X_2\textcolor{blue}{\pm}Y}\right]\,.
\end{alignat}
These formulae exactly recover the derivative weight-shifting operators derived in~\cite{Arkani-Hamed:2018kmz} as we demonstrate in Appendix~\ref{app:derivativews}. We now move on to the derivation of these weight-shifting matrices, following the same logic as was used in Section~\ref{subsec: contact diagram}.

\begin{eBox3}
\textbf{Derivation:}\\ 
\textit{Shifting $\nu$:}
     We start off by writing out the master integrals for this diagram, with a shifted weight,
    \ba
    \label{eq: shifting xi inset}\vec I_{\alpha_1,\alpha_2,\nu+1}=  \int  \frac{\ud\eta_1 e^{iX_1\eta_1}}{(-\eta_1)^{2+\alpha_1+\dev}}\frac{\ud\eta_2 e^{iX_2\eta_2}}{(-\eta_2)^{2+\alpha_2+\dev}}\begin{bmatrix}
       G^{++}_{\nu+1}(Y,\eta_1,\eta_2)\\G^{+-}_{\nu+1}(Y,\eta_1,\eta_2)\\G^{-+}_{\nu+1}(Y,\eta_1,\eta_2)\\G^{--}_{\nu+1}(Y,\eta_1,\eta_2)\\i(-\eta_1)^{2\xi+3}\delta_{12}
   \end{bmatrix},
    \ea
    where we have introduced the shorthand notation $\delta_{ij}=\delta(\eta_i-\eta_j)$. Recall that the bulk-to-bulk propagators $G_{\nu}^{\pm\pm}$, defined in \reef{eqn: G definition}, are entirely in terms of the $h^\pm_\nu$ and $\bar{h}^\pm_\nu$. This means that the shift relations \reef{eqn: h shift relations} can be directly applied to $G_\nu^{\pm\pm}$. At the level of the vector $\vec{I}_{\alpha_1,\alpha_2,\nu}$, these become:
    \begin{equation}\label{eqn: single exchange first shift example}
    \begin{split}
         \vec I_{\alpha_1,\alpha_2,\nu+1}&=-D\cdot \vec I_{\alpha_1,\alpha_2,\nu}  \;+ \frac{\nu^2}{Y^2}E_0 \cdot\vec I_{\alpha_1+1,\alpha_2+1,\nu}+\frac{i\nu}{Y}E_2\cdot \vec I_{\alpha_1+1,\alpha_2,\nu}+\frac{i\nu}{Y}E_1\cdot \vec I_{\alpha_1,\alpha_2+1,\nu}\,,
         \end{split}
    \end{equation}
     where we have defined the following matrices:

\be\label{eq:DE matrix}
D= \text{{\small$\begin{bmatrix}
        1 & 0 & 0 & 0 & 0 \\
        0& -1& 0 & 0 & 0 \\ 
        0&0& -1 & 0 & 0\\
         0&0& 0&1& 0\\
         0& 0& 0& 0& -1
    \end{bmatrix}$}} ,\ 
E_1= \text{{\small$\begin{bmatrix}
        1 & 1 & 0 & 0 & 0 \\
        1& 1& 0 & 0 & 0 \\ 
        0&0& -1 & -1 & 0\\
         0&0& -1&-1& 0\\
         0& 0& 0& 0& 0
    \end{bmatrix}$}},\ 
    E_2= \text{{\small$\begin{bmatrix}
        1 & 0 & 1 & 0 & 0 \\
         0&-1& 0&-1& 0\\
        1 & 0 & 1 & 0 & 0 \\
         0&-1& 0&-1& 0\\
         0& 0& 0& 0& 0
    \end{bmatrix}$}},\
    E_0=\text{{\small$\begin{bmatrix}
        1 & 1 & 1 & 1 & 0 \\
         1&1& 1&1& 0\\
        1 & 1 & 1 & 1 & 0 \\
         1&1& 1&1& 0\\
         0& 0& 0& 0& 0
    \end{bmatrix}$}}\,.
\ee
Just as in the case of the contact diagram, we consider two different ways of transforming this into a relation between just two different vectors. Both of these methods employ the $A$-matrix associated with the single-exchange diagram.
\begin{enumerate}
        \item \emph{Equal $\alpha$:} Through the shift relation \reef{eqn: introducing alpha shift}, we see that we can convert all the vectors on the right side of \reef{eqn: single exchange first shift example} $\vec{I}_{\alpha_1,\alpha_2,\nu}$ as follows,
\ba\label{eqn: equal alpha single-exchange shift}
\vec I_{\alpha_1,\alpha_2,\nu+1}=& \left(-D +  \frac{\nu}{Y}E_2\cdot \left[A^{(X_1)}_{\alpha_1+1,\alpha_2}\right]^{-1}+ \frac{\nu}{Y} E_1\cdot \left[A^{(X_2)}_{\alpha_1,\alpha_2+1}\right]^{-1}\right.\\
&\left.\;-\frac{\nu^2}{Y^2} E_0\cdot \left[A^{(X_1)}_{\alpha_1+1,\alpha_2}\cdot A^{(X_2)}_{\alpha_1+1,\alpha_2+1}\right]^{-1}\right)\cdot \vec I_{\alpha_1,\alpha_2,\nu}\,.
\ea
    
    \item \emph{Shifted $\alpha$}: This time we use the shift relation \reef{eqn: introducing alpha shift} to convert all the vectors on the right side of \reef{eqn: single exchange first shift example} to $\vec{I}_{\alpha_1+1,\alpha_2+1,\nu}$ as follows,
\ba\label{eqn: shifted alpha single exchange example}
\vec I_{\alpha_1,\alpha_2,\nu+1}=& \left(D\cdot A^{(X_1)}_{\alpha_1+1,\alpha_2}\cdot A^{(X_2)}_{\alpha_1+1,\alpha_2+1} - \frac{\nu}{Y}E_2\cdot A^{(X_2)}_{\alpha_1+1,\alpha_2+1}\right.\\
&\left.\;-  \frac{\nu}{Y}E_1\cdot A^{(X_1)}_{\alpha_1+1,\alpha_2+1}+ \frac{\nu^2}{Y^2}E_0\right)\cdot \vec I_{\alpha_1+1,\alpha_2+1,\nu}\,.
\ea
    \end{enumerate}
Using the algorithm defined in \cite{massivekinflow}, we can construct this diagram's $A$-matrix and its $\partial_{X_1}$- and $\partial_{X_2}$-components. Substituting these and their inverses into~\reef{eqn: equal alpha single-exchange shift} and \reef{eqn: shifted alpha single exchange example}, we arrive at the expressions \reef{eq: M for exchange} and \reef{eqn: curly M for exchange} for $M$ and $\mathcal{M}$, respectively.

\end{eBox3}

\subsection{Kronecker Product Representation}\label{subsec: kronecker notation}

The examples above can be generalized succinctly using Kronecker products. In this section, we introduce Kronecker products (see for example \cite[Chapter 4.2]{kronecker} for a standard reference), quote  some of their important properties and demonstrate how they can be used to condense all the master integrals associated with an arbitrary tree-level diagram into a single expression. 

\vskip4pt

The Kronecker product of an $m\times n$ matrix $A=(a_{ij})_{i\in[m], \ j\in[n]}$ and a $p\times q$ matrix $B$ is defined to be the $mp\times nq$ matrix 
\begin{equation}
    A\otimes B = \begin{bmatrix}
        a_{11}B & a_{12}B & \cdots & a_{1n}B \\
        a_{21}B & a_{22}B & \cdots & a_{2n}B \\
        \vdots & \vdots & \ddots & \vdots \\
        a_{m1}B & a_{m2}B & \cdots & a_{mn}B
    \end{bmatrix}.
\end{equation}
In the context of this paper, important features of Kronecker products to keep in mind are:
\begin{itemize}
    \item As an operation, the Kronecker product is associative, but not commutative.
    \item As a map, the Kronecker product is bilinear: For an $m\times n$ matrix $A$, $p\times q$ matrices $B$ and $C$ and a scalar $r$ in the relevant field, we have that
    \ba
        A\otimes(B+C)&=A\otimes B+A\otimes C\,,\\
        (B+C)\otimes A&=B\otimes A+C\otimes A\,,\\
        A\otimes(rB)&=(rA)\otimes B=r(A\otimes B)\,.
    \ea
    \item  Kronecker products satisfy the following \textit{mixed product property}: Consider two sequences of $\ell$ matrices $\{A_i\}_{i\in[\ell]}$ and $\{B_i\}_{i\in[\ell]}$, where $A_i$ has dimensions $m_i\times n_i$ and $B_i$ has dimensions $n_i\times q_i$. Then we have the identity,
    \begin{equation}\label{eq:mpp}
        \left[\bigotimes_{i=1}^\ell A_i\right] \cdot\left[\bigotimes_{i=1}^\ell B_i\right]=\bigotimes_{i=1}^\ell(A_i\cdot B_i)\,,
    \end{equation}
    where $(\,\cdot\,)$ denotes matrix multiplication.
\end{itemize}
Using $\mathbbm{I}_n$ to denote an $n\times n$ identity matrix, note that $\mathbbm{I}_{n}^{\otimes m}=\mathbbm{I}_{n^m}$. Consequently, in what follows we will often write chains of Kronecker products of identity matrices as larger identity matrices.

\vskip4pt

\begin{figure}[t!]
\centering
\adjustbox{valign=t}{\includegraphics[page=5]{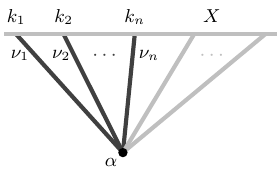}}
\quad\quad\quad\quad
\adjustbox{valign=t}{\includegraphics[page=6]{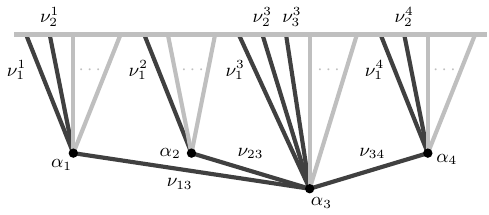}}
\caption{On the left is an arbitrary contact diagram, with $n$ arbitrary mass lines (dark grey) and any number of conformally coupled lines (light grey). On the right is an example of a tree-level diagram with internal lines. The kinematic variables have been suppressed to avoid unnecessary clutter.}
\label{fig: notation examples}
\end{figure}

Now consider a contact diagram with $n$ arbitrary mass lines, each labelled with a weight, $\nu_i=\half+\xi_i$, and energy, $k_i$, plus any number of conformally coupled lines, with total energy $X$. See the left side of Figure~\ref{fig: notation examples} for a visual representation of this diagram. Using Kronecker products, we can write the vector $\vec{I}$ of master integrals corresponding to this diagram in the form
\begin{equation}\label{eqn: general contact integral vector}
    \vec{I}_{\alpha,\bm{\nu}}=\int\frac{\ud\eta}{(-\eta)^{1+\alpha}}e^{i X\eta}\bigotimes_{i=1}^n\frac{1}{(-\eta)^{\xi_i}}\begin{bmatrix}
        h_{\nu_i}^+(k_i,\eta)\\h_{\nu_i}^-(k_i,\eta)
    \end{bmatrix},
\end{equation}
where $\bm{\nu}=(\nu_1,\ldots,\nu_n)$ is the tuple of $\nu_i$ indices relevant to the diagram, and $\alpha$ is the single vertex parameter. In what follows, we will represent the shift $\nu_i\rightarrow\nu_i+1$ at the level of the tuple $\bm{\nu}$ by writing $\bm{\nu}\rightarrow\bm{\nu}+\bm{e}_i$, where $\bm{e}_i$ is the $i$th standard unit vector.

\vskip4pt

More generally, we can construct such a vector for an arbitrary tree-level diagram. Suppose the diagram has $n$ vertices, vertex $i$ having $m_i$ arbitrary mass bulk-to-boundary lines. Label each of these by a weight $\nu_{j}^i=\half+\xi^i_{j}$ and external energy $k^i_j$, where $j\in[m_i]$. As in the case of the contact diagram, each of the vertices $i$ can have any number of additional conformally coupled bulk-to-boundary lines attached to it. At vertex $i$, call the combined external energy associated with these $X_i$ and call the vertex parameter $\alpha_i$. Denote by $L$ the set of pairs $(i,j)$, with $i<j$, labelling the vertices of the diagram that share a bulk-to-bulk propagator. Impose the following total order $\prec$ on $L$: $(i,j)\succ(p,q)$ when $i>p$ and $(i,j)\succ(i,q)$ when $j>q$. Suppose $(i,j)\in L$, then label the momentum exchanged by the corresponding bulk-to-bulk propagator by $Y_{ij}$ and the weight of the particle it corresponds to by $\nu_{ij}=\half+\xi_{ij}$. A visual example of such a diagram with the associated labelling conventions can be seen on the right side of Figure~\ref{fig: notation examples}. Using this notation, the vector of master integrals corresponding to an arbitrary tree-level diagram can be written in the form
\begin{equation}\label{eqn: general tree-level integral vector}
   \begin{split}
   \vec{I}_{\bm{\alpha},\bm{\nu}}=\int\prod_{i=1}^n&\left[\frac{\ud\eta_i}{(-\eta_i)^{1+\alpha_i}}e^{i\eta_iX_i}\right]\left[\bigotimes_{i=1}^n\otimes_{j=1}^{m_i}\frac{1}{(-\eta_i)^{\xi_j^i}}\begin{bmatrix}h^+_{\nu^i_j}(k^i_j,\eta_i)\\h^-_{\nu^i_j}(k^i_j,\eta_i)\end{bmatrix}\right]\
   \\
   &\otimes\bigotimes_{(i,j)\in L}\frac{1}{(\eta_i\eta_{j})^{\xi_{ij}}}\begin{bmatrix}
       G^{++}_{\nu_{ij}}(Y_{ij},\eta_i,\eta_{j})\\G^{+-}_{\nu_{ij}}(Y_{ij},\eta_i,\eta_{j})\\G^{-+}_{\nu_{ij}}(Y_{ij},\eta_i,\eta_{j})\\G^{--}_{\nu_{ij}}(Y_{ij},\eta_i,\eta_{j})\\i(-\eta_i)^{2\xi_{ij}+1}\delta_{ij}
   \end{bmatrix},
   \end{split}
\end{equation}
where the order in which the Kronecker product over $L$ is taken matches the total order $\prec$ defined on $L$ above, starting with the smallest pair $(i,j)$ and ending with the largest pair. To avoid clutter, we have introduced the short form notation
\ba\label{eq: vec alpha nu 0}
    \bm{\alpha}&=(\alpha_1,\ldots,\alpha_n)\,,\\
    \bm{\nu}&=(\nu_1^1,\ldots,\nu_{m_1}^{1},\ldots,\nu_1^n,\ldots,\nu_{m_n}^{n},\{\nu_{ij}:(i,j)\in L\})\,.
\ea
To avoid introducing unnecessary complications, rather than specifying an order on the entries of $\bm{\nu}$, we simply define $\bm{e}^i_j$ and $\bm{e}_{ij}$ to be the standard unit vectors with $1$ in the appropriate position to realise the shifts in $\nu^i_j\rightarrow\nu^i_j+1$ and $\nu_{ij}\rightarrow\nu_{ij}+1$, respectively, in $\bm{\nu}$.

\vskip4pt

\noindent For example, for the diagram on the right of Figure~\ref{fig: notation examples}, we have
    \ba\label{eq: vec alpha nu}
    \bm{\alpha}&=(\alpha_1,\alpha_2,\alpha_3,\alpha_4)\,,\\
        \bm{\nu}&=(\nu^1_1,\nu^1_2,\nu^2_1,\nu^3_1,\nu^3_2,\nu_3^3,\nu^4_1,\nu^4_2,\nu_{13},\nu_{23},\nu_{34})\,.
    \ea
     We realise shifts in the parameters at the level of the tuples by, for example,
    \ba\label{eq: vec alpha nu2}
        \bm{\alpha}+\bm{e}_3&=(\alpha_1,\alpha_2,\alpha_3+1,\alpha_4)\,,\\
        \bm{\nu}+\bm{e}^3_2&=(\nu^1_1,\nu^1_2,\nu^2_1,\nu^3_1,\nu^3_2+1,\nu^3_3,\nu^4_1,\nu^4_2,\nu_{13},\nu_{23},\nu_{34})\,,\\
        \bm{\nu}+\bm{e}_{13}&=(\nu^1_1,\nu^1_2,\nu^2_1,\nu^3_1,\nu^3_2,\nu^3_3,\nu^4_1,\nu^4_2,\nu_{13}+1,\nu_{23},\nu_{34})\,.
    \ea
Note that the vector of master integrals $\vec{I}_{\bm{\alpha},\bm{\nu}}$ corresponding to an arbitrary tree-level diagram has $2^N5^{|L|}$ entries, $N$ being the total number of arbitrary mass external lines. The Kronecker product notation has allowed us to write this vector in two lines, and in such a way that one can still do algebraic manipulations of it through the properties of the Kronecker product.

\vskip4pt

As mentioned above, Kronecker products do not commute. Implicit in the expressions \reef{eqn: general contact integral vector} and \reef{eqn: general tree-level integral vector} is a choice of ordering of the vertices and bulk-to-boundary propagators. For the results presented in this paper, the particular choice of ordering does not matter, so long as one remains consistent with this choice.

\subsection{General Diagram}\label{subsec: generic shifting}
So far, we have discussed how the weight-shifting matrix $M$ is derived for two simple examples.
In this section, we use the Kronecker product representation to compactly generalize these results to arbitrary tree-level diagrams. We show how the mixed product property of Kronecker products allows us to consider local transformations within a given vector of master integrals. Using this, we derive the weight-shifting matrix for a shift in weight corresponding to both external lines and internal lines. 
\subsubsection*{External Shift---Contact Diagram}
\label{subsubsec:extshift}

To start, we illustrate our method in the case of a contact diagram with $n$ arbitrary mass external lines, each labelled by a weight $\nu_i$. This is the diagram on the left in Figure~\ref{fig: notation examples}, whose master integrals are the entries of the vector $\vec{I}_{\alpha,\bm{\nu}}$ in \reef{eqn: general contact integral vector}. First, notice that the shift relations \reef{eqn: h shift relations} in $h_{\nu}^{\pm}$ and $\bar{h}_{\nu}^\pm$ can be organised into the following matrix equation,
\ba\label{eqn: shift relations}
    \frac{1}{(-\eta)^{\xi+1}}\begin{bmatrix}h^+_{\nu+1}(k,\eta)\\h^-_{\nu+1}(k,\eta)\end{bmatrix}&=\frac{1}{(-\eta)^\xi}\left(iB-\frac{\nu}{k\eta}C\right) \cdot\begin{bmatrix}h^+_{\nu}(k,\eta)\\h^-_{\nu}(k,\eta)\end{bmatrix},
\ea
where the matrices $B$ and $C$ are defined as in \reef{eq: BC def}. Using the mixed product property of Kronecker products, we can isolate the action of this shift to a single mass-parameter $\nu_j$ in $\vec{I}_{\alpha,\bm{\nu}}$. Explicitly, we \emph{pad} the matrices $B$ and $C$ with additional $2\times 2$ identity matrices, defining the following two $2^n\times2^n$ matrices, 
\ba\label{eq: Bj Cj}
\left\{\begin{array}{lr}B_j\\C_j\end{array}\right\}&=\mathbbm{I}_{2^{j-1}}\otimes \left\{\begin{array}{lr}B\\C\end{array}\right\} \otimes\mathbbm{I}_{2^{n-j}}\,.
\ea
Note that the dimension of these square matrices matches the number of elements in the vector $\vec{I}_{\alpha,\bm{\nu}}$. This ensures that matrix multiplication with $\vec{I}_{\alpha,\bm{\nu}}$ is always possible and that it will return a vector with the same number of entries.

\vskip4pt

\noindent\textit{Shifting $\nu$:}
Using \reef{eqn: shift relations} and the mixed product property of Kronecker products \reef{eq:mpp}, we find
\begin{equation}\label{eqn: generic shifting 1}
\vec{I}_{\alpha,\bm{\nu}+\bm{e}_j}=iB_j\cdot\vec{I}_{\alpha,\bm{\nu}}+\frac{\nu_j}{k_j}C_j \cdot\vec{I}_{\alpha+1,\bm{\nu}}\,.
\end{equation}
This is the generalization of \reef{eq: intermediate I} to an arbitrary contact diagram. An explicit example that works through this in the case where $n=2$ can be found in Appendix~\ref{app: n=2 contact nu shift example}.

\vskip4pt

\noindent\textit{Shifting $\alpha$:} 
As described in the example of Section~\ref{subsec: contact diagram}, there are two types of weight-shifting operators we can derive. We called these $\mathcal{M}$ and $M$, where $\mathcal{M}$ carries out a shift in both $\alpha$ and $\nu$, whilst $M$ shifts just $\nu$. We now generalise these matrices to an arbitrary contact diagram. Recall from Section~\ref{subsubsec: A-matrix alpha shift} that $A$-matrices can be used to shift $\alpha$ as follows,
\begin{equation}
    -i\vec{I}_{\alpha,\bm{\nu}}=A^{(X)}_{\alpha+1}\cdot\vec{I}_{\alpha+1,\bm{\nu}}\,.
\end{equation}
Using this, as well as~\reef{eqn: generic shifting 1}, we find that the matrices $\mathcal{M}^{\mathrm{ext}}_j$ and $M^{\mathrm{ext}}_j$, which perform the shifts
\begin{equation}
    \quad\vec{I}_{\alpha,\bm{\nu}+\bm{e}_j}=\mathcal{M}^{\text{ext}}_j \cdot \vec{I}_{\alpha+1,\bm{\nu}}\,
    \quad \text{and}
\quad\vec{I}_{\alpha,\bm{\nu}+\bm{e}_j}=M^{\text{ext}}_j \cdot \vec{I}_{\alpha,\bm{\nu}}\,,
\end{equation}
are given by
\begin{equation}\label{eq: external generic}
    \mathcal{M}^{\mathrm{ext}}_j  =-B_j\cdot A^{(X)}_{\alpha+1}+\frac{\nu_j}{k_j}C_j\, \quad \text{and} \quad M^{\mathrm{ext}}_j=iB_j-i\frac{\nu_j}{k_j}C_j[A^{(X)}_{\alpha+1}]^{-1}\,,
\end{equation}
where we have adorned these weight-shifting matrices with $\mathrm{ext}$ and $j$ to emphasize that they carry out a shift in the weight of the $j$th external line of the contact diagram.

\vskip4pt

In Appendix~\ref{app: derivation of contact weight-shifting}, we construct $C_j[A^{(X)}_{\alpha+1}]^{-1}$ for an arbitrary contact diagram and thus present an explicit, closed form formula for $M^{\mathrm{ext}}_j$. 

\vskip4pt

\subsubsection*{External Shift---General Diagram} The generalization of the discussion above to shifting the mass of an external line in a generic tree-level\footnote{All our results apply equally to loop integrands. However, describing the weight-shifting operators for such diagrams requires introducing labels that can distinguish between multiple edges connecting the same vertices. We believe that implementing appropriate labelling would only serve to confuse the overall message of this section and so we keep all our discussions at tree-level.} diagram is straightforward. We will mainly be adapting the notation such that the matrices act on the vector $\vec{I}_{\bm{\alpha},\bm{\nu}}$ from \reef{eqn: general tree-level integral vector}. The analogue of the matrices \reef{eq: Bj Cj} is
\ba\label{eq: Bj Cj p}
\left\{\begin{array}{lr}B^i_{j}\\C^i_{j}\end{array}\right\}&=\mathbbm{I}_{2^{\sum_{\ell=1}^{i-1}m_\ell}}\otimes\mathbbm{I}_{2^{j-1}}\otimes \left\{\begin{array}{lr}B\\C\end{array}\right\} \otimes\mathbbm{I}_{2^{m_i-j+\sum_{\ell=i+1}^nm_\ell}}\otimes\mathbbm{I}_{5^{|L|}}\,,
\ea
where $\mathbbm{I}_{5^{|L|}}$ corresponds to the part of the Kronecker product in~\reef{eqn: general tree-level integral vector} that is responsible for internal lines. In keeping with the conventions defined in Section~\ref{subsec: kronecker notation}, the upper $i$ and lower $j$ notation is used to specify that $B$ and $C$ act on the $j$th external line of the $i$th vertex. Just as in the case of the contact diagram, the shift in the weight of the $j$th external line of the $i$th vertex is then given by 
\begin{equation}
 \vec{I}_{\bm{\alpha},\bm{\nu}+\bm{e}^i_j}=iB^i_{j} \cdot \vec{I}_{\bm{\alpha},\bm{\nu}}+\frac{\nu^i_j}{k^i_j}C^i_{j} \cdot \vec{I}_{\bm{\alpha}+\bm{e}_i,\bm{\nu}}\,.
\end{equation}
Using the $A$-matrix component that performs the shift
\ba\label{eq: shift alpha generic}
-i\vec{I}_{\bm{\alpha},\bm{\nu}}&=A_{\bm{\alpha}+\bm{e}_i}^{(X_i)}\cdot\vec{I}_{\bm{\alpha}+\bm{e}_i,\bm{\nu}}\,,
\ea
we can construct the analogue of $\mathcal{M}^{\mathrm{ext}}_j$ and $M^{\mathrm{ext}}_j$, generalised to an arbitrary tree-level diagram
\begin{equation}
    \vec{I}_{\bm{\alpha},\bm{\nu}+\bm{e}^i_j}= \mathcal{M}^{i\,\text{ext}}_{j}\cdot \vec{I}_{\bm{\alpha}+\bm{e}_i,\bm{\nu}} \quad \text{and} \quad \vec{I}_{\bm{\alpha},\bm{\nu}+\bm{e}^i_j}= {M}^{i\,\text{ext}}_{j}\cdot \vec{I}_{\bm{\alpha},\bm{\nu}}\,,
\end{equation}
where 
\begin{equation}
    \mathcal{M}^{i\,\text{ext}}_{j}=-B^i_{j}A^{(X_i)}_{\bm{\alpha}+\bm{e}_i}+\frac{\nu^i_j}{k^i_j}C^i_{j}\, \quad \text{and} \quad {M}^{i\,\text{ext}}_{j}=iB^i_{j} -i\frac{\nu^i_j}{k^i_j}C^i_{j} \left[A^{(X_i)}_{\bm{\alpha}+\bm{e}_i}\right]^{-1}\,.
\end{equation}
From these equations we can see that all the additional complexity in shifting the masses of arbitrary external edges comes from the increase in complexity of taking derivatives, as represented by the $A$-matrix.

%%%%%%%%%%%%%%%%%%%%%%%%%%%%%%%%%%%%%%%%%%%%%%%%
\subsubsection*{Internal Shift}
It remains to consider shifts in the weights associated with internal lines. In what follows, we will be constructing weight-shifting matrices that realise this shift for an arbitrary tree-level diagram. As such, we will be working with matrices that act on the vector $\vec{I}_{\bm{\alpha},\bm{\nu}}$ of \reef{eqn: general tree-level integral vector}. This section can be read as a generalisation of the two-site exchange example shown in Section~\ref{subsec: exchange example} to an arbitrary tree-level diagram.

\vskip4pt

First, note that we can use the identity \reef{eqn: h shift relations} to derive the following shift relations between the functions $G^{\pm\pm}_\nu$ defined in \reef{eqn: G definition},
\ba\label{eqn: shift relations exchange}
    \frac{1}{(\eta_i\eta_j)^{\xi+1}}\begin{bmatrix}G^{++}_{\nu+1}(Y,\eta_i,\eta_j)\\G^{+-}_{\nu+1}(Y,\eta_i,\eta_j)\\G^{-+}_{\nu+1}(Y,\eta_i,\eta_j)\\G^{--}_{\nu+1}(Y,\eta_i,\eta_j)\\i(-\eta_i)^{2\xi+3}\delta_{ij}\end{bmatrix}&=-\frac{1}{(\eta_i\eta_j)^{\xi}}\left[D+\frac{i\nu}{Y}\left(\frac{E_2}{\eta_i}+\frac{E_1}{\eta_j}+\frac{i\nu E_0}{Y\eta_i\eta_j}\right)\right]\cdot\begin{bmatrix}G^{++}_{\nu}(Y,\eta_i,\eta_j)\\G^{+-}_{\nu}(Y,\eta_i,\eta_j)\\G^{-+}_{\nu}(Y,\eta_i,\eta_j)\\G^{--}_{\nu}(Y,\eta_i,\eta_j)\\i(-\eta_i)^{2\xi+1}\delta_{ij}\end{bmatrix},
\ea
where we have used the $5\times5$ matrices $D$, $E_0$, $E_1$ and $E_2$ that were introduced in \reef{eq:DE matrix}. Following the same logic used in the construction of \reef{eq: Bj Cj} and \reef{eq: Bj Cj p}, we pad with identity matrices so that we can isolate the action of $D$, $E_0$, $E_1$ and $E_2$ on specific components of $\vec{I}_{\bm{\alpha},\bm{\nu}}$. Explicitly, we define
\ba
\left\{\begin{array}{lr}D_{ij}\\E_{ij0}\\E_{ij1}\\E_{ij2}\\\end{array}\right\}&=\mathbbm{I}_{2^{\sum_{k=1}^nm_k}}\otimes\left[\bigotimes_{\substack{(k,l)\in L \\ %|
(k,l)\prec(i,j)}}\mathbbm{I}_{5}\right]\otimes\left\{\begin{array}{lr}D\\E_{0}\\E_{1}\\E_{2}\\\end{array}\right\}\otimes\left[\bigotimes_{\substack{(k,l)\in L \\% |
(k,l)\succ(i,j)}}\mathbbm{I}_{5}\right].
\ea
By the mixed product property of Kronecker products, these matrices act non-trivially on the component of $\vec{I}_{\bm{\alpha},\bm{\nu}}$ corresponding to the internal line with weight $\nu_{ij}$, and leave all other components unchanged. Using these matrices, the relation \reef{eqn: single exchange first shift example} from the two-site exchange example generalises to
\begin{equation}\label{eqn: generic shifting 3}
\vec{I}_{\bm{\alpha},\bm{\nu}+\bm{e}_{ij}}=-D_{ij}\cdot\vec{I}_{\bm{\alpha},\bm{\nu}}+\frac{\nu_{ij}^2}{Y_{ij}^2}E_{ij0}\cdot\vec{I}_{\bm{\alpha}+\bm{e}_i+\bm{e}_j,\bm{\nu}}+i\frac{\nu_{ij}}{Y_{ij}}E_{ij1}\cdot\vec{I}_{\bm{\alpha}+\bm{e}_j,\bm{\nu}}+i\frac{\nu_{ij}}{Y_{ij}}E_{ij2}\cdot\vec{I}_{\bm{\alpha}+\bm{e}_i,\bm{\nu}}\,.
\end{equation}
As before, we can use the relevant $A$-matrix component to perform the shift written in \reef{eq: shift alpha generic}. Using this and the relation above, we can construct the matrices $\mathcal{M}^{\mathrm{int}}_{ij}$ and $M^{\mathrm{int}}_{ij}$ which realise the shifts
\ba
    \vec{I}_{\bm{\alpha},\bm{\nu}+\bm{e}_{ij}}=\mathcal{M}^{\mathrm{int}}_{ij}\cdot\vec{I}_{\bm{\alpha}+\bm{e}_i+\bm{e}_j,\bm{\nu}}\, \quad \text{and} \quad \vec{I}_{\bm{\alpha},\bm{\nu}+\bm{e}_{ij}}=M^{\mathrm{int}}_{ij}\cdot\vec{I}_{\bm{\alpha},\bm{\nu}}\,.
\ea
Explicitly, these are
\ba\label{eq:ExplicitM}
\mathcal{M}^{\mathrm{int}}_{ij}&=-\frac{\nu_{ij}}{Y_{ij}}\big(E_{ij1}\cdot A^{(X_{i})}_{\bm{\alpha}+\bm{e}_i+\bm{e}_j}+E_{ij2}\cdot A^{(X_j)}_{\bm{\alpha}+\bm{e}_i+\bm{e}_j}-\frac{\nu_{ij}}{Y_{ij}}E_{ij0}\big)+D_{ij}\cdot A^{(X_{i})}_{\bm{\alpha}+\bm{e}_i}\cdot A^{(X_{j})}_{\bm{\alpha}+\bm{e}_i+\bm{e}_j}\,,\\
M^{\mathrm{int}}_{ij}&=-D_{ij}+\frac{\nu_{ij}}{Y_{ij}}(E_{ij2}-\frac{\nu_{ij}}{Y_{ij}}E_{ij0}[A^{(X_j)}_{\bm{\alpha}+\bm{e}_i+\bm{e}_j}]^{-1})\cdot[A^{(X_i)}_{\bm{\alpha}+\bm{e}_i}]^{-1}+\frac{\nu_{ij}}{Y_{ij}}E_{ij1}[A^{(X_j)}_{\bm{\alpha}+\bm{e}_j}]^{-1}\,.
\ea
We conclude this section by mentioning that 
computing the $\mathcal{M}$-matrices above is relatively straightforward. This is because, using the diagrammatic rules introduced in \cite{massivekinflow}, $A$-matrices are easily constructed for arbitrary tree-level diagrams. On the other hand, calculating $M$-matrices is more complicated since it always involves inverting the relevant component of the $A$-matrix. As mentioned before, in the particular case of an arbitrary contact diagram, we were able to derive a closed formula for $M^{\mathrm{ext}}_j$---see Appendix~\ref{app: derivation of contact weight-shifting}. Whilst we believe that similar expressions can be found for $M^{i \ \mathrm{ext}}_{j}$ and $M^{\mathrm{int}}_{ij}$, the inversion of the relevant $A$-matrix components for these diagrams is less straightforward. One way to see this is to note that the inversion of these matrices is `non-local' in the sense that it involves every part of the associated diagram, not just the part being shifted or its immediate neighbours. 

\vskip4pt

We would like to briefly mention a potentially surprising omission for those familiar with kinematic flow: the absence of diagrammatic rules for this process in our work. 
In practice, these diagrammatic rules become more involved than the kinematic flow rules themselves, so they no longer serve their main purpose: organising the underlying combinatorial structure. This can be seen in the case of exchange diagram in \reef{eqn: curly M for exchange} and \reef{eq: M for exchange} where despite the apparent combinatorial structure, there is mixing between all the other master integrals with several possible ``letters'' that makes a diagrammatical representation less demonstrative. 
It is straightforward to derive flow rules for the weight-shifting matrices from the kinematic flow rules of the $A$-matrix, using matrix multiplication, addition, and inversion (for the $M$-matrix) via~\reef{eq:ExplicitM} and~\reef{eq: external generic}. However, the resulting diagrammatic rules are neither practical nor illuminating.
One way to derive such diagrammatic rules is 
through construction of weight-shifting matrices based on the kinematic flow rules of the $A$-matrix  as a starting point.  
For this reason, while, in principle, they exist,  we do not include such diagrammatic rules in our analysis.

%%%%%%%%%%%%%%%%%%%%%%%%%%%%%%%%%%%%%
\section{Massless from Conformally Coupled}
\label{sec: massless}
Central to modern cosmology is the computation of de Sitter 
correlation functions involving massless fields. This focus on massless fields is because they yield scale-invariant power spectra that match observations of the Cosmic Microwave Background~\cite{Planck:2019nip,pajer2024field}. They are therefore the only fields we have observational evidence for and are good candidates for slow-roll inflation~\cite{Liddle:1994dx}. In addition, de Sitter correlators of fields with non-zero mass are exponentially suppressed at late times by the expansion of the universe. We therefore predict that the only measurable correlators today are those of only massless scalar fields (and gravitons). This does not mean that we should only compute 
wavefunction coefficients involving solely massless fields---the internal lines may be of arbitrary mass. Indeed, the entire cosmological collider research program~\cite{Arkani-Hamed:2015bza,Jazayeri:2022kjy,Reece:2022soh,Chen:2022vzh,Hubisz:2024xnj,Green:2026yev,Sohn:2024xzd,Chakraborty:2023qbp} is based around evaluating diagrams involving massive internal lines, as these serve as signatures for the existence of massive particles during inflation.

\vskip4pt

Inflationary spacetime is typically modelled as dS$_4$, which, in our conventions, corresponds to the case $d=3$. As described in the introduction, it is common within the de Sitter literature to compute correlation functions for conformally coupled external fields due to their simplicity, particularly when using the differential equations approach. In dS$_4$ these fields, with $\nu=1/2$, are conveniently exactly an integer shift away from being massless, $\nu=3/2$. Therefore, perhaps the most important cosmological application of weight-shifting operators has been in computing massless wavefunction coefficients
from conformally coupled seeds~\cite{Baumann:2019oyu,Arkani-Hamed:2018kmz}. In this section, we apply our weight-shifting matrices to do just this. Where our approach differs from earlier work is in the simplicity of implementation described in Section~\ref{subsec: generic shifting}. In particular, our weight-shifting matrices don't require taking derivatives and act graph-locally. As we will show, this simplification enables us to compute wavefunction coefficients
involving massless fields that were inaccessible using previous methods. It is likewise straightforward to act repeatedly with these matrices to compute the wavefunction coefficients of exceptional mass scalars from this same conformally coupled seed. We discuss this further in Appendix~\ref{app:Repeated Shift}.
\newpage
\vskip4pt

Before moving on, we note that the shift $\nu=1/2\rightarrow\nu=3/2$ is, in some sense, the most trivial example of the shifts encoded by the relation \eqref{eqn: h shift relations}. This can be seen explicitly by expanding the Hankel functions in the bulk-to-boundary propagators at $\nu=1/2$ and $\nu=3/2$,
\begin{equation}\label{eqn: similarity of massless and conf coup}
    g_{\nu=\half}(\eta)= e^{ik \eta} \,\longrightarrow\, g_{\nu=1+\half}(\eta)= \frac{1-i k \eta}{k}e^{ik \eta}.
\end{equation}
Thus, the shift $\nu=1/2\rightarrow\nu=3/2$ in a given diagram can be understood as a sum (up to multiplication by constants) over two copies of that diagram, the second having a shifted vertex parameter. However, to systematically perform this sum for arbitrary diagrams would, in practice, require keeping track of all the momentum-dependent coefficients. This will be done for us by the weight-shifting matrices. 

\vskip4pt

Throughout this section, we will use dark grey dotted lines to represent massless fields and light grey, solid lines for conformally coupled ones.

\subsection{Single Massless Contact}\label{sec:masslesscontact}
We start by revisiting the simplest example: shifting the weight of a single external edge in a contact diagram. The weight-shifting matrix $M$ for this example was first computed in Section~\ref{subsec: contact diagram}. Specialising to $\nu=1/2$ (or equivalently $\xi=0$), we find
\ba\begin{bmatrix}
        \psi_{\alpha,\frac{3}{2}}^{+}\\
        \psi_{\alpha,\frac{3}{2}}^-
    \end{bmatrix}=
    M\cdot\begin{bmatrix}
        \psi_{\alpha,\frac{1}{2}}^{+}\\
        \psi_{\alpha,\frac{1}{2}}^-
    \end{bmatrix}=i\begin{bmatrix}
        1-\frac{X+k}{2k(1+\alpha)}&-\frac{X-k}{2k(1+\alpha)}\\
        -\frac{X+k}{2k(1+\alpha)}&-1-\frac{X-k}{2k(1+\alpha)}
    \end{bmatrix}\cdot\begin{bmatrix}
        \psi_{\alpha,\frac{1}{2}}\\
        0
    \end{bmatrix}=i\begin{bmatrix}
        1-\frac{k+X}{2k(1+\alpha)}\\
        -\frac{k+X}{2k(1+\alpha)}
    \end{bmatrix}\psi_{\alpha,\frac{1}{2}}\,,
\ea
where $\psi_{\alpha,\frac{1}{2}}$ corresponds to the contact diagram with all conformally coupled external lines.
Since we have set $d=3$, we can also fix $\alpha=3-n$, where $n$ is total number of lines attached to the vertex, using~\eqref{eqn: physical alpha}. Consequently, adding the two entries above yields the wavefunction coefficient corresponding to a contact diagram with $n-1$ conformally coupled lines and one massless line,
\ba\label{eq:massless1shift}
    \psi_{\alpha,\frac{3}{2}} \;=\;
    \adjustbox{valign=c,scale=1}{\includegraphics[page=7]{img/TikzPictures.pdf}}\;=\;
i\frac{k\alpha-X}{k(1+\alpha)}\psi_{\alpha,\frac{1}{2}}\,.
\ea
Alternatively, we can derive $\psi_{\alpha,\frac{3}{2}}$ using the weight-shifting matrix that also shifts $\alpha$
\ba
    \begin{bmatrix}
        \psi_{\alpha,\frac{3}{2}}^+\\\psi_{\alpha,\frac{3}{2}}^-
    \end{bmatrix}=
    \mathcal{M}\cdot \begin{bmatrix}
        \psi_{\alpha+1,\frac{1}{2}}^+\\\psi_{\alpha+1,\frac{1}{2}}^-
    \end{bmatrix}=\begin{bmatrix}
        -\frac{1+\alpha}{X+k}+\frac{1}{2k}&\frac{1}{2k}\\\frac{1}{2k}&\frac{1+\alpha}{X-k}+\frac{1}{2k}
    \end{bmatrix}\cdot\begin{bmatrix}
        \psi_{\alpha+1,\frac{1}{2}}\\0
    \end{bmatrix}=\begin{bmatrix}
        \frac{1}{2k}-\frac{1+\alpha}{k+X}\\\frac{1}{2k}
    \end{bmatrix}\psi_{\alpha+1,\frac{1}{2}}.
\ea
As before, we can add the two entries together. This yields
\ba\label{eqn: massless contact with alpha shift}
    \psi_{\alpha,\frac{3}{2}}=\frac{X-\alpha k}{k(X+k)}\psi_{\alpha+1,\frac{1}{2}}\,.
\ea
Given the simplicity of this example, we can confirm the results \eqref{eq:massless1shift} and \eqref{eqn: massless contact with alpha shift} by performing the integrals prescribed by the Feynman rules directly. We have that
\ba\label{eq:CCintegral}
    \psi_{\alpha,\frac{1}{2}}&=\int_{-\infty}^0 \frac{\ud\eta}{(-\eta)^{1+\alpha}}e^{i(X+k)\eta}=i^\alpha(k+X)^{\alpha}\,\Gamma(-\alpha)\,,\\
    \psi_{\alpha,\frac{3}{2}}&=\frac{1}{k} \int_{-\infty}^0 \frac{\ud\eta}{(-\eta)^{2+\alpha}}\,(1-i k \eta)e^{i(X+k)\eta}=i^{\alpha+1}(k+X)^{\alpha}\left(\frac{X}{k}-\alpha\right)\Gamma(-\alpha-1)\,,
\ea
from which the relationship \eqref{eq:massless1shift} can easily be validated. The same integrals can be used to validate \eqref{eqn: massless contact with alpha shift} by replacing $\alpha$ with $\alpha+1$ on both sides.

\subsection{All Massless Contact}\label{sec:allmassless}

A slightly more involved example is the four-point contact diagram with all massless lines. Whilst this diagram is still computable using conventional methods, we use it to demonstrate how our weight-shifting matrices can be applied recursively. A pictorial representation of this diagram is given below
\ba
    \psi_{\alpha,\mathbf{\frac{3}{2}}}=\adjustbox{valign=c,scale=1}{\includegraphics[page=8]{img/TikzPictures.pdf}},
\ea
where $\mathbf{\frac{3}{2}}=\left(\frac{3}{2},\frac{3}{2},\frac{3}{2},\frac{3}{2}\right)$ is the shorthand notation introduced in \reef{eq: vec alpha nu 0} indicating all external lines are massless. In this section, we compute this diagram by repeatedly applying weight-shifting matrices to the four-point contact diagram with all conformally coupled lines. These matrices are all $16\times 16$ which is too large to be helpfully represented. Therefore, we will not write them explicitly, instead focusing on how they should be used. As in the previous example, we first consider the case where $\alpha$ is kept un-shifted. In this case, the vector of massless master integrals is given by
\ba\label{eqn: 4-point massless integral}
    \vec{I}_{\alpha,\mathbf{\frac{3}{2}}}={M}_{\alpha,\frac{1}{2},\frac{3}{2},\frac{3}{2},\frac{3}{2}}\cdot {M}_{\alpha,\frac{1}{2},\frac{1}{2},\frac{3}{2},\frac{3}{2}}\cdot {M}_{\alpha,\frac{1}{2},\frac{1}{2},\frac{1}{2},\frac{3}{2}}\cdot {M}_{\alpha,\frac{1}{2},\frac{1}{2},\frac{1}{2},\frac{1}{2}}\cdot \vec{I}_{\alpha,\mathbf{\frac{1}{2}}}\,.
\ea
The wavefunction coefficient $\psi_{\alpha,\mathbf{\frac{3}{2}}}$ is then the sum of the entries of $\vec{I}_{\alpha,\mathbf{\frac{3}{2}}}$ and is given by
\ba\label{eqn: 4-point massless}
    \psi_{\alpha,\mathbf{\frac{3}{2}}}=
        \frac{e_4(1+\alpha)(2+\alpha)(4+\alpha)-k_Te_3(2+\alpha)(4+\alpha)+k_T^2e_2(4+\alpha)-k_T^4}{e_4(1+\alpha)(2+\alpha)(4+\alpha)}\psi_{\alpha,\mathbf{\frac{1}{2}}}\,,
\ea
where we have introduced the elementary symmetric polynomials,
\ba
    k_T&=k_1+k_2+k_3+k_4\,,& e_2&=k_1k_2+k_1k_3+k_1k_4+k_2k_3+k_2k_4+k_3k_4\,,&\\e_4&=k_1k_2k_3k_4\,,&e_3&=k_1k_2k_3+k_1k_2k_4+k_1k_3k_4+k_2k_3k_4\,.
\ea
Note that, as a function, $\psi_{\alpha,\bm{\frac{1}{2}}}$ is identical to $\psi_{\alpha,\frac{1}{2}}$ from the previous section. The difference between them is in interpretation. In $\psi_{\alpha,\frac{1}{2}}$, $k$ is the energy of the line being shifted, whilst $X$ represents the combined energy of the $n-1$ conformally coupled lines not being shifted. In $\psi_{\alpha,\bm{\frac{1}{2}}}$, all four of the lines with energies $k_1,\, k_2,\, k_3$ and $k_4$ are shifted, and $X$ is set to zero.

\vskip4pt

So far, we have been treating $\alpha$ as a parameter. As explained in Section~\ref{subsec: setting and notation}, we need to set $\alpha$ to the value prescribed by~\eqref{eqn: physical alpha} to obtain the physically-relevant version of $\psi_{\alpha,\mathbf{\frac{3}{2}}}$. Since this diagram has four lines, in $d=3$ we need to set $\alpha=-1$. We immediately run into the problem that \eqref{eqn: 4-point massless} has a pole at $\alpha=-1$. To investigate the nature of this pole, we compute the same vector $\vec{I}_{\alpha,\bm{\frac{3}{2}}}$, except this time using the weight-shifting matrices $\mathcal{M}$ that realise a shift in both weight and vertex parameter. This is done below
\ba\label{eq:masslessalphashift}
    \vec{I}_{\alpha,\mathbf{\frac{3}{2}}}=\mathcal{M}_{\alpha,\frac{1}{2},\frac{3}{2},\frac{3}{2},\frac{3}{2}}\cdot \mathcal{M}_{\alpha+1,\frac{1}{2},\frac{1}{2},\frac{3}{2},\frac{3}{2}}\cdot \mathcal{M}_{\alpha+2,\frac{1}{2},\frac{1}{2},\frac{1}{2},\frac{3}{2}}\cdot \mathcal{M}_{\alpha+3,\frac{1}{2},\frac{1}{2},\frac{1}{2},\frac{1}{2}}\cdot \vec{I}_{\alpha+4,\mathbf{\frac{1}{2}}}\,.
\ea
The vector $\vec{I}_{\alpha+4,\bm{\frac{1}{2}}}$ contains a single non-zero entry, $\psi_{\alpha+4,\bm{\frac{1}{2}}}$. Using~\eqref{eq:CCintegral} and taking the limit as $\alpha\rightarrow -1$, we find that
\ba\label{eqn: log with scale}
    \lim_{\alpha\rightarrow-1}\psi_{\alpha+4,\mathbf{\frac{1}{2}}}=\lim_{\alpha\rightarrow -1}(ik_T)^{\alpha+4}\Gamma(-\alpha-4)=\lim_{\alpha\rightarrow -1}\frac{ik_T^3}{36}\left[11-6\left(\gamma+\log(ik_T)-\frac{1}{\alpha+1}\right)\right],
\ea
where $\gamma$ is the Euler--Mascheroni constant. This shows that $\psi_{\alpha+4,\bm{\frac{1}{2}}}$ has a simple pole at $\alpha=-1$. The standard interpretation of such a singularity is that we should cut off the integral at some finite time $\eta_0$ rather than allowing our universe to continue inflating into the asymptotic future~\cite{DanielCosmo}. Introducing such a cut-off and setting $\alpha$ to $-1$, this integral becomes
\ba\label{eqn: additional integral}
    \lim_{\eta_0\rightarrow 0}\psi_{3,\mathbf{\frac{1}{2}}}(\eta_0)&=\lim_{\eta_0\rightarrow 0}\int_{-\infty}^{\eta_0}\frac{\ud\eta}{(-\eta)^{4}}e^{ik_T\eta}\\&=\lim_{\eta_0\rightarrow 0}\left(-\frac{1}{3\eta_0^3}-\frac{ik_T}{2\eta_0^2}+\frac{k_T^2}{2\eta_0}+\frac{ik_T^3}{36}\left[11-6(\gamma+\log(-ik_T\eta_0))\right]\right).
\ea
We see that it is thus possible to exchange the simple pole at $\alpha=-1$ for a logarithmic time divergence. Indeed, we could have argued this without performing the additional integral \eqref{eqn: additional integral}. Notice that the pole at $\alpha=-1$ in~\eqref{eqn: log with scale} comes alongside a logarithm that has scale. A regulated version of \eqref{eqn: additional integral} must therefore introduce some additional factor in the logarithm to make it scale invariant, as we see in \eqref{eqn: additional integral}. There are some additional polynomial divergences in \eqref{eqn: additional integral} that are less clearly connected to $\alpha$. These can be regulated away through the introduction of counterterms in the action, a process called holographic renormalisation~\cite{Skenderis:2002wp}.

\vskip4pt

Mollifying the pole at $\alpha=-1$ by introducing the cut-off $\eta_0$, we find the sum of the entries in $\vec{I}_{\alpha,\bm{\frac{3}{2}}}$ computed using \eqref{eq:masslessalphashift} to be
\begin{align}\nonumber
\psi_{-1,\mathbf{\frac{3}{2}}}(\eta_0)&=
        \lim_{\alpha\rightarrow-1}\left(\frac{2(3k_Te_2-k_T^3-3e_3)}{k_T^3e_4}-\frac{(k_T^4-5k_T^2e_2+11k_Te_3-6e_4)(1+\alpha)}{k_T^4e_4}\right)\psi_{\alpha+4,\mathbf{\frac{1}{2}}}\\\label{eqn: 4-point all massless 1}&=i\frac{k_T^3-3k_Te_2+3e_3}{3e_4}\left(\log(-i k_T\eta_0)+\gamma\right)-i\frac{4k_T^4-9k_T^2e_2+9e_4}{9k_Te_4}\,.
\end{align}
Up to polynomial time divergences, this agrees with the result arrived at by directly applying the Feynman rules and performing the time integrals
\ba\label{eqn: 4-point all massless 2}
    \lim_{\eta_0\rightarrow 0}&\psi_{-1,\mathbf{\frac{3}{2}}}(\eta_0)=\lim_{\eta_0\rightarrow 0}\frac{1}{e_4}\int_{-\infty}^{\eta_0}\ud\eta(-\eta)^{-4}(1-ik_1\eta)(1-ik_2\eta)(1-ik_3\eta)(1-ik_4\eta)e^{ik_T\eta}\\&=i\frac{k_T^3-3k_Te_2+3e_3}{3e_4}\left(\gamma+\log(-ik_T\eta_0)\right)-i\frac{4k_T^4-9k_T^2e_2+9e_4}{9e_4k_T}-\frac{k_T^2-2e_2}{2e_4\eta_0}-\frac{1}{3e_4\eta_0^3}\,.
\ea

\vskip4pt

Returning to the expression for $\psi_{\alpha,\bm{\frac{3}{2}}}$ derived using \eqref{eqn: 4-point massless integral}, we can perform an identical analysis. We find that $\psi_{\alpha,\bm{\frac{1}{2}}}$ is finite in the limit as $\alpha\rightarrow-1$, meaning that the simple pole at $\alpha=-1$ is introduced by the weight-shifting matrices $M$
\ba\label{eq:allmasslessshift}
    \psi_{-1,\mathbf{\frac{3}{2}}}&=
        i\lim_{\alpha\rightarrow -1}\left(\frac{k_T(k_T^3-3k_Te_2+3e_3)}{3e_4(\alpha+1)}-\frac{4k_T^4-9k_T^2e_2+9e_4}{9e_4}\right)\frac{1+(\alpha+1)(\gamma+\log(ik_T))}{k_T}\,,
\ea
where the factor outside the brackets has come from expanding $\psi_{-1,\mathbf{\frac{1}{2}}}$ to subleading order in the $\alpha\rightarrow -1$ limit. Performing the same trick of replacing the pole at $\alpha=-1$ with a logarithmic time divergence, this agrees exactly with~\eqref{eqn: 4-point all massless 1}.

\subsection{Massless Internal Line}

In both of the previous examples, the conformally coupled solutions are rational in the external kinematics and, even when the integrals diverge they can be computed easily. In this section, we compute the wavefunction coefficient 
for four conformally coupled particles exchanging a massless field: 
\ba
    \psi_{\alpha_1,\alpha_2,\frac{3}{2}}=\adjustbox{valign=c}{\includegraphics[page=9]{img/TikzPictures.pdf}}
\ea
We do this by acting with a single weight-shifting matrix on the seed-diagram whose lines are all conformally coupled. At the level of the master integrals associated with each of these diagrams, this is the equation
\begin{equation}\label{eqn: massless exchange example}
    \vec{I}_{\alpha_1,\alpha_2,\frac{3}{2}}=M_{\alpha_1,\alpha_2,\frac{1}{2}}\cdot\vec{I}_{\alpha_1,\alpha_2,\frac{1}{2}}\,,
\end{equation}
where $M$ is the matrix \eqref{eq: M for exchange} constructed in Section~\ref{subsec: exchange example}, with the weight $\nu$ of the internal line in the diagram being shifted set to $1/2$ (or equivalently $\xi=0$). Since for this diagram the physically-relevant values for $\alpha_1$ and $\alpha_2$ are the same, henceforth we set $\alpha\equiv\alpha_1=\alpha_2$. In doing this, the matrix $M$ is given by
\ba\label{eqn: exchange matrix applied}
    M_{\alpha,\alpha,\frac{1}{2}}=-{\small\frac{1}{4Y^2(1+\alpha)^2}\begin{bmatrix}
        X_1^{(-\alpha)}X_2^{(-\alpha)}& X_2^{(-)}X_1^{(-\alpha)}& X_1^{(-)}X_2^{(-\alpha)}& X_2^{(-)}X_1^{(-)}& 2(X_1+X_2)Y\\ 
        X_2^{(-)}X_1^{(-\alpha)}& X_2^{(\alpha)}X_1^{(-\alpha)}& X_1^{(-)}X_2^{(+)}& X_1^{(-)}X_2^{(\alpha)}& 2(X_1+X_2)Y\\ X_1^{(+)}X_2^{(-\alpha)}&X_2^{(-)}X_1^{(+)}& X_1^{(\alpha)}X_2^{(-\alpha)}& X_2^{(-)}X_1^{(\alpha)}& 2(X_1+X_2)Y\\ X_2^{(+)}X_1^{(+)}& X_1^{(+)}X_2^{(\alpha)}&X_2^{(+)}X_1^{(\alpha)}& X_2^{(\alpha)}X_1^{(\alpha)}& 2(X_1+X_2)Y\\0&0&0&0& -4Y^2(1+\alpha)^2
    \end{bmatrix}},
\ea
where we have defined $X_i^{(\pm\alpha)}=X_i\pm(1+2\alpha)Y,\, X_i^{(\pm)}=X_i^{(\pm 0)}$ for notational compactness.

\vskip4pt

The conformally coupled master integrals are given by~\cite{Baumann:2025qjx}
\ba\label{eq:conformal exchange}
    \vec{I}_{\alpha,\alpha,\frac{1}{2}}=i^{2\alpha}\Gamma(-2\alpha)\begin{bmatrix}
        0&F^{12}_\alpha&F^{21}_\alpha&0&(X_1+X_2)^{2\alpha}
    \end{bmatrix}^{\text{T}}\,,
\ea
where the prefactor has been derived by considering the explicit time integrals, and we have introduced the notation:
\ba
    F^{ij}_\alpha=-\frac{1}{\alpha}(X_i+Y)^{2\alpha}
\; {}_2F_1\bigg[\begin{array}{c}
-\alpha \,,\, -2\alpha\\[-1pt]
1-\alpha
\end{array}\bigg\rvert \,
\frac{Y-X_j}{X_i+Y}
\,\bigg]\,.
\ea
Substituting this vector and the matrix \eqref{eqn: exchange matrix applied} into \eqref{eqn: massless exchange example} and rearranging, we find
\ba
    \vec{I}_{\alpha,\alpha,\frac{3}{2}}=-\frac{i^{2\alpha}\Gamma(-2\alpha)}{4Y^2(1+\alpha)^2}\begin{bmatrix}
        2(X_1+X_2)^{1+2\alpha}Y+X_2^{(-)}X_1^{(-\alpha)}F_{\alpha}^{12}+X_1^{(-)} X_2^{(-\alpha)}F_{\alpha}^{21}\\2(X_1+X_2)^{1+2\alpha}Y+X_2^{(\alpha)} X_1^{(-\alpha)}F_{\alpha}^{12}+X_1^{(-)} X_2^{(+)}F_{\alpha}^{21}\\2(X_1+X_2)^{1+2\alpha}Y+X_1^{(+)}X_2^{(-)}F_{\alpha}^{12}+X_1^{(\alpha)} X_2^{(-\alpha)}F_{\alpha}^{21}\\2(X_1+X_2)^{1+2\alpha}Y+X_1^{(+)}X_2^{(\alpha)}F_{\alpha}^{12}+X_1^{(\alpha)} X_2^{(+)}F_{\alpha}^{21}\\-4(1+\alpha)^2Y^2(X_1+X_2)^{2\alpha}
    \end{bmatrix}.
\ea
Recall from Section \ref{subsec: massive kinematic flow} that the function associated with a given Feynman diagram is the sum of all the non-collapsed master integrals of that diagram. The vector $\vec{I}_{\alpha,\alpha,\frac{3}{2}}$ contains a single collapsed integral, the final entry. The wavefunction coefficient $\psi_{\alpha,\alpha,\frac{3}{2}}$ is thus the sum of the first four entries of $\vec{I}_{\alpha,\alpha,\frac{3}{2}}$, and is given by
\ba
    \psi_{\alpha,\alpha,\frac{3}{2}}&=-\frac{i^{2\alpha}\Gamma(-2\alpha)}{Y^2(1+\alpha)^2}\left[(X_1-Y\alpha)(X_2+Y\alpha)F_\alpha^{12}+(1\leftrightarrow 2)\right]-\frac{2i^{2\alpha}\Gamma(-2\alpha)(X_1+X_2)^{1+2\alpha}}{Y(1+\alpha)^2}\,.
\ea
We remind the reader that, in line with our conventions, this is actually the time-ordered contribution to the wavefunction coefficient. To recover the full wavefunction coefficient we simply add the product of lower point functions that gives us the disconnected component,
\ba
    \psi_{\alpha,\alpha,\frac{3}{2}}^{\text{D}}&=-\psi_{\alpha,\frac{3}{2}}(X_1,Y)\psi_{\alpha,\frac{3}{2}}(X_2,Y)\\&=\frac{i^{2\alpha}\Gamma(-\alpha)^2}{Y^2(1+\alpha)^2}((X_1+Y)(X_2+Y))^{\alpha}\left(X_1-Y\alpha\right)\left(X_2-Y\alpha\right)\,.
\ea
Thus, the full wavefunction\footnote{It is also possible to recover this expression from a single shift operation by considering the vector
\ba
    \vec{\mathcal{I}}_{\alpha,\alpha,\frac{1}{2}}=i^{2\alpha}\begin{bmatrix}
        \Gamma(-\alpha)^2(X_1+Y)^\alpha(X_2+Y)^\alpha~&~\Gamma(-2\alpha)F^{12}_\alpha~&~\Gamma(-2\alpha)F^{21}_\alpha~&~0~&~\Gamma(-2\alpha)(X_1+X_2)^{2\alpha}
    \end{bmatrix}^{\text{T}}\,.
\ea
This is because, whether or not we include the disconnected contribution, the master integrals satisfy the same differential equation and are thus subject to the same shift relations.} coefficient is
\ba 
    \psi_{\alpha,\alpha,\frac{3}{2}}&+\psi_{\alpha,\alpha,\frac{3}{2}}^{\text{D}}=\frac{i^{2\alpha}\Gamma(-2\alpha)}{Y^2(1+\alpha)^2}\left[\frac{\Gamma(-\alpha)^2}{\Gamma(-2\alpha)}((X_1+Y)(X_2+Y))^\alpha(X_1-Y\alpha)(X_2-Y\alpha)\right.\\&\left.-(X_2+Y\alpha)(X_1-Y\alpha)F_\alpha^{12}-(X_1+Y\alpha)(X_2-Y\alpha)F_\alpha^{21}-2Y(X_1+X_2)^{1+2\alpha}\right]\,.
\ea
From this we see the comment made beneath~\eqref{eqn: similarity of massless and conf coup} in action---namely that that the complexity of massless diagrams is essentially the same as that of conformally coupled diagrams. Had we followed the differential equations method introduced in~\cite{massivekinflow}, we would find that the differential equations governing massless contributions have the same level of complexity as arbitrary mass contributions. Only when one specialises to conformally coupled solutions do the differential equations simplify. Nonetheless, the ability to express the massless contributions as a simple linear combinations of the conformally coupled ones ensures that the massless diagrams are no more complex.

\vskip4pt

Before closing this section, we note that the results presented here can be compared with the results from~\cite[Section 4.3]{Arkani-Hamed:2018kmz}. In~\cite{Arkani-Hamed:2018kmz}, the authors use a more involved differential operator based approach. We review this approach in Appendix~\ref{app:derivativews}, and demonstrate the equivalence of both methods when applied to this example.

\subsection{The Cosmological Collider}
So far, all the examples we carried out were illustrative and have reproduced known results. We now turn to a non-trivial and phenomenologically significant example: a single exchange diagram in which all external lines are massless and the internal line has arbitrary mass,
\ba\label{eqn: massless arbitrary mass exchange}
    \psi_{\alpha_1,\alpha_2,\mathbf{\frac{3}{2}},\nu}=\adjustbox{valign=c}{\includegraphics[page=10]{img/TikzPictures.pdf}}\,.
\ea

\vskip4pt

For the same reasons as in the previous example, henceforth we set $\alpha\equiv\alpha_1=\alpha_2$. Using our weight-shifting matrices, {\small$\psi_{\alpha,\alpha,\mathbf{\frac{3}{2}},\nu}$} can be computed as the sum of the non-collapsed entries in {\small$\vec{I}_{\alpha,\alpha,\bm{\frac{3}{2}},\nu}$}. The vector {\small$\vec{I}_{\alpha,\alpha,\bm{\frac{3}{2}},\nu}$} itself can be computed from the conformally coupled seed vector {\small$\vec{I}_{\alpha,\alpha,\bm{\frac{1}{2}},\nu}$} by recursively applying weight-shifting matrices $M$ to shift the four external lines from conformally coupled to massless. Note that {\small$\vec{I}_{\alpha,\alpha,\bm{\frac{1}{2}},\nu}$} contains five non-zero entries, {\small$\psi^{\pm\pm}_{\alpha,\alpha,\bm{\frac{1}{2}},\nu}$} and the collapsed component {\small$J_{\alpha,\alpha,\bm{\frac{1}{2}}}$}. The $80$ entries of {\small$\vec{I}_{\alpha,\alpha,\bm{\frac{3}{2}},\nu}$} are then linear combinations of these---the precise linear combinations encoded by the product of four weight-shifting matrices. These weight-shifting matrices can be derived straightforwardly using the formulae derived in Section~\ref{subsec: generic shifting}, the relevant $A$-matrices being similarly simple to construct using the kinematic flow rules of \cite{massivekinflow}. 

\vskip4pt

Rather than present the full system of $80$ master integrals, we simply quote the sum of all the non-collapsed entries, that being the wavefunction coefficient $\psi_{\alpha,\alpha,\mathbf{\frac{3}{2}},\nu}$. We find
\ba\label{eqn: arbitrary exchange massless legs 2}
    \psi_{\alpha,\alpha,\mathbf{\frac{3}{2}},\nu}=\frac{1}{e_4(1+\alpha-\xi)^2(2+\alpha+\xi)^2}\left[\sum_{a,b=\pm}\mathcal{F}_1^a\mathcal{F}_2^b\psi_{\alpha,\alpha,\mathbf{\frac{1}{2}},\nu}^{ab}-\frac{2k_T Y\mathcal{G}J_{\alpha,\alpha,\mathbf{\frac{1}{2}}}}{(1+2\alpha)(3+2\alpha)}\right]\,,
\ea
where we have defined
\ba
    \mathcal{F}_i^\pm&=X_i^{(\pm)}( X_i(1+\alpha-\xi)\mp Y)-k_{2i-1}k_{2i}(1+\alpha-\xi)(2+\alpha+\xi)\,,\\
    \mathcal{G}&=-Y^2(3+2\alpha)+k_T^2(1+\alpha-\xi)(2+\alpha+\xi)-e_2(3+2\alpha)(1+\alpha-\xi)(2+\alpha+\xi)\\&+X_1X_2(1+2\alpha)(3+2\alpha)(1+\alpha-\xi)\,.
\ea
We present the equivalent relationship derived using $\mathcal{M}$ in Appendix~\ref{app:Massless Alt} as, despite the computational simplicity, the final expressions are substantially more complex. This result could, in principle, have been computed using the differential operator based approach of~\cite{Arkani-Hamed:2018kmz}. However, in using weight-shifting matrices, we have reduced the repeated action of differential operators to simple matrix multiplication, enabling us to write explicit formulae rather than leaving the computation of these objects implicit. At an operational level we have avoided repeatedly invoking the derivative identities satisfied by the hypergeometric functions from which $\psi^{\pm\pm}_{\alpha,\alpha,\bm{\frac{1}{2}},\nu}$ is defined.

\vskip4pt

The functions $\psi^{\pm\pm}_{\alpha,\alpha,\mathbf{\frac{1}{2}},\nu}$ are left implicit because they have not, themselves, been computed from the kinematic flow. However, following the methods outlined in~\cite{Arkani-Hamed:2018kmz} it should be possible to solve for them as some hypergeometric terms plus an infinite sum coming from the presence of the source term $J$. To give at least one concrete expression, we can consider massless particles exchanging a conformally coupled field. In this case
$\vec{I}_{\alpha,\alpha,\mathbf{\frac{1}{2}},\frac{1}{2}}$ is equal to~\eqref{eq:conformal exchange},
\begin{align}
    \psi^{++}_{\alpha,\alpha,\mathbf{\frac{1}{2}},\frac{1}{2}}=\psi^{--}_{\alpha,\alpha,\mathbf{\frac{1}{2}},\frac{1}{2}}=0\,,\quad\psi^{+-}_{\alpha,\alpha,\mathbf{\frac{1}{2}},\frac{1}{2}}=C F_\alpha^{12}\,,\quad\psi^{-+}_{\alpha,\alpha,\mathbf{\frac{1}{2}},\frac{1}{2}}=C F_\alpha^{21}\,,\quad J_{\alpha,\alpha,\mathbf{\frac{1}{2}}}=C k_T^{2\alpha}\,,
\end{align}
where $C=i^{2\alpha}\Gamma(-2\alpha)$. Going to dS$_4$ by taking the limit $\alpha\rightarrow 0$ is a singular limit as the massless field has non-shift symmetric interactions. 
The master integrals in this limit, corresponding to the exchange of a massive field with mass parameter $\nu=\half+\xi$, were computed in \cite{massivekinflow} up to order $O(\xi)$.
In the next section, we discuss how to relate this to finite interactions that involve derivative couplings.

%========================================
\subsection{Derivative Interactions}
Our weight-shifting matrix approach reveals the following interesting feature of all-massless (or exceptional mass, discussed in Appendix~\ref{app:Repeated Shift}) tree-level diagrams: their $2^N5^{|L|}$ master integrals can be written entirely in terms of the $3^{|L|}$ master integrals associated with the same diagram but with all conformally coupled weights. This discrepancy is particularly stark in the contact case, in which a single conformally coupled integral generates all $2^n$ massless contributions.
One could take this as a sign that uplifting the problem to a set of master integrals is introducing a large amount of redundancy. In this section, we argue against this by highlighting a feature of these master integrals that has yet to be acknowledged: they can be repurposed as master integrals for the same diagram, but where any number of the polynomial interactions have been replaced with derivative interactions. To see this explicitly, first note that the mode function $\phi_\nu$ that satisfies the free-field Euler--Lagrange equation is related to the rescaled mode function $g_\nu$ through \cite{massivekinflow}
\be
\phi_\nu(k,\eta)= \mathcal{N}\,(-\eta)^{\frac{d}{2}-\nu} g_\nu(k,\eta)\,, \quad 
\text{with} \quad \mathcal{N} =\frac{iH^{\frac{d-1}{2}}}{\sqrt{2k}}e^{-i\frac{\pi\nu}{2}}\,.
\ee
Now recall from Section~\ref{sec: master integrals} that the basic building blocks of the master integrals are the functions $h_\nu^\pm$, and that $g_\nu$ is the sum of these. Thus
\ba\label{eq:Derivs}
    \phi_\nu(k,\eta)&=\mathcal{N}(-\eta)^{\frac{d}{2}-\nu}\left(h^+_\nu(k,\eta)+h^-_\nu(k,\eta)\right)\,,\\
    \eta \partial_\eta \phi_\nu(k,\eta)&=\mathcal{N}(-\eta)^{\frac{d}{2}-\nu}\left[ik\eta\left(h^+_\nu(k,\eta)-h^-_\nu(k,\eta)\right)+\left(\frac{d}{2}-\nu\right)\left(h^+_\nu(k,\eta)+h^-_\nu(k,\eta)\right)\right]\,,\\\eta \partial_i \phi_\nu(k,\eta)&=\mathcal{N}(-\eta)^{\frac{d}{2}-\nu}ik_i\eta\left(h^+_\nu(k,\eta)+h^-_\nu(k,\eta)\right)\,.
\ea
We now demonstrate how these relations can be used to compute the contact diagram~\eqref{eq:massless1shift} from Section~\ref{sec:masslesscontact}, but where we replace some of the propagators with various derivatives thereof. We denote the result of such a computation by $\psi_{\alpha,\nu}(\,\cdot\,)$, where the item in the brackets specifies the interaction vertex under consideration. As before, $\vec{I}_{\alpha,\nu}$ represents the vector of master integrals associated with the diagram when the interaction vertex is simply polynomial. Defining the vectors $\vec{v}_\pm\equiv[1 \ \ \pm1 ]$, the wavefunction coefficients for simple derivative interactions are
\ba\label{eqn: shifting interaction}
    \psi_{\alpha,\nu}(\phi_{\frac{1}{2}}^{n-1}\phi_\nu)&=\vec{v}_+\cdot \vec{I}_{\alpha,\nu}\,,\\
    \psi_{\alpha,\nu}(\phi_{\frac{1}{2}}^{n-1}\eta\partial_\eta\phi_\nu)&=\left(-k\vec{v}_-\cdot A^{(X)}_{\alpha,\nu}+\left(\frac{d}{2}-\nu\right)\vec{v}_+\right)\cdot \vec{I}_{\alpha,\nu}\,,\\
    \psi_{\alpha,\nu}(\phi_{\frac{1}{2}}^{n-2}\eta^2\partial_i\phi_{\frac{1}{2}}\partial^i\phi_\nu)&=k_n^2\vec{v}_+\cdot \vec{I}_{\alpha-2,\nu}=k_n^2\vec{v}_+\cdot A^{(X)}_{\alpha-1,\nu}\cdot A^{(X)}_{\alpha,\nu}\cdot \vec{I}_{\alpha,\nu}\,.
\ea
The first line is the standard interpretation of the master integrals used throughout the previous sections. In the final line we have symmetrised over the $n-1$ conformally coupled fields and used momentum conservation to write the coefficient in terms of the energy of the field $\phi_\nu$, $k_n$.

\vskip4pt

Again, using the mixed product properties of Kronecker products, just like in Section~\ref{subsec: generic shifting} we can appropriately pad with identity matrices to uplift the shifts in interaction type \eqref{eqn: shifting interaction} to arbitrary bulk-to-boundary propagators in arbitrary tree-level diagrams. When working with bulk-to-bulk propagators one needs to be slightly more careful since~\cite{Abolhasani_2022}
\ba
    \partial_{\eta}\partial_{\eta'}G_{\nu,F}(k,\eta,\eta')=\langle 0\rvert T(\partial_\eta \phi_\nu(k,\eta)\partial_{\eta'}\phi_\nu(k,\eta'))\lvert 0\rangle +\frac{i}{(-H\eta)^{d-1}} \delta(\eta-\eta')\,,
\ea
where $|0\rangle$ denotes the Bunch--Davies vacuum of the free-theory. Thus, diagrams involving bulk-to-bulk propagators will also pick up contributions from collapsed terms.

\vskip4pt

As described in earlier sections, the most phenomenologically interesting examples are those involving primarily massless fields. In $d$ spatial dimensions, this corresponds to the case $\nu=d/2$. In this case, the middle relation in \eqref{eqn: shifting interaction} simplifies substantially
\ba
    \psi_{\alpha,\frac{d}{2}}(\phi_{\frac{1}{2}}^{n-1}\eta\partial_\eta\phi_{\frac{d}{2}})&=-k\vec{v}_-\cdot A^{(X)}_{\alpha,\frac{d}{2}}\cdot \vec{I}_{\alpha,\frac{d}{2}}\,.
\ea
Specialising to $d=3$, and using that $\vec{I}_{\alpha,\frac{3}{2}}$ can be re-written as the action of the appropriate weight-shifting matrix on the conformally coupled case $\vec{I}_{\alpha,\frac{1}{2}}$, we find
\ba\label{eq:deta}
    \psi_{\alpha,\frac{3}{2}}(\phi_{\frac{1}{2}}^{n-1}\eta\partial_\eta\phi_{\frac{3}{2}})&=-k\vec{v}_-\cdot A^{(X)}_{\alpha,\frac{3}{2}}\cdot M_{\alpha,\frac{1}{2}}\cdot \vec I_{\alpha,\frac{1}{2}}=
        \frac{i\alpha k}{X+k}\psi_{\alpha,\frac{1}{2}}=k\psi_{\alpha-1,\frac{1}{2}}\,,
\ea
where the final equality just represents the identity
\ba\label{eq:ccderiv}
    \eta\partial_\eta \phi_{\frac{3}{2}}=ik\eta \phi_{\frac{1}{2}}\,.
\ea
Derivative interactions for massless fields can thus be interpreted as shifts in $\alpha$ from the conformally coupled case.

\vskip4pt

To further illustrate this point, we return to the example from Section~\ref{sec:allmassless} except this time consider interactions involving time derivatives. We find
\ba
    \psi_{-1,\bm{\frac{3}{2}}}(({\eta}\partial_{\eta}\phi_{\frac{3}{2}})^4)&=\lim_{\alpha\rightarrow -1}\frac{e_4\alpha(\alpha-1)(\alpha-2)(\alpha-3)}{k_T^4}\psi_{\alpha,\bm{\frac{1}{2}}}=-\frac{24e_4}{k_T^5}\,,\\
    \psi_{-1,\bm{\frac{3}{2}}}(\phi_{\frac{3}{2}}({\eta}\partial_{\eta}\phi_{\frac{3}{2}})^3)&=\lim_{\alpha\rightarrow -1}\frac{\alpha(\alpha-1)(e_3e_4(\alpha-2)-e_3^2k_T+2e_2e_4k_T)}{4e_4k_T^3}\psi_{\alpha,\bm{\frac{1}{2}}}\\&=\frac{3e_3 e_4+e_3^2k_T-2e_2e_4k_T}{2e_4k_T^4}\,,\\\psi_{-1,\bm{\frac{3}{2}}}(\phi_{\frac{3}{2}}^2({\eta}\partial_{\eta}\phi_{\frac{3}{2}})^2)&=\lim_{\alpha\rightarrow -1}\frac{e_2^2k_T^2-e_2\alpha(e_4+e_3k_T-e_4\alpha)+k_T^2(e_4(2+3\alpha)-2e_3k_T)}{6e_4k_T^2}\psi_{\alpha,\bm{\frac{1}{2}}}\\&=-\frac{e_2^2k_T^2+2e_2e_4+e_2e_3k_T-k_T^2e_4-2e_3k_T^3}{6e_4k_T^3}\,.
\ea
We have presented these as a limits to provide the expression for arbitrary $\alpha$ alongside the physically-relevant result. We have also used~\eqref{eq:CCintegral} to give us $\psi_{-1,\bm{\frac{1}{2}}}=-k_T^{-1}$. Note that for fewer than two derivatives the result diverges. Nonetheless, we can employ the same technique as was used in Section~\ref{sec:allmassless} to replace this divergence with a logarithmic time divergence
\begin{align}\nonumber
    \psi_{-1,\bm{\frac{3}{2}}}(\phi^3_{\frac{3}{2}}\eta\partial_{\eta}\phi_{\frac{3}{2}})&=\lim_{\alpha\rightarrow -1}-\frac{k_T^4-e_2k_T^2(4+\alpha)+e_3k_T(2+\alpha)(4+\alpha)-e_4(1+\alpha)(2+\alpha)(4+\alpha)}{e_4(1+\alpha)(2+\alpha)}\\&=\frac{k_Te_3+2e_2k_T^2-k_T^4-3e_4}{4k_Te_4}+\frac{3e_3-3e_2k_T+k_T^3}{4e_4}(\gamma+\log(-ik_T\eta_0))\,.
\end{align}
Thus, we have computed all four-point contact interactions involving first-order time derivatives of massless fields from a single conformally coupled contact diagram.

Combining these results with those from Section~\ref{sec:allmassless} we can recover an exact and explicit result in de Sitter for massless external scalars. To maintain some of the interesting features of the massless modefunctions we consider a spatial derivative interaction (time derivatives look like conformally coupled wavefunctions with a shifted $\alpha$~\eqref{eq:deta}),
\begin{align}
    &\psi_{0,0,\mathbf{\frac{3}{2}},\frac{3}{2}}(\eta^2\phi \partial_i\phi\partial^i \phi)=-\frac{1}{4}(Y^2+k_1^2+k_2^2)(Y^2+k_3^2+k_4^2)\psi_{-2,-2,\mathbf{\frac{3}{2}},\frac{3}{2}}\\\nonumber&=\frac{(Y^2+k_1^2+k_2^2)(Y^2+k_3^2+k_4^2)}{e_4k_T^3Y}\left[\vphantom{\frac{X_2^{(+)}}{{X_2^{(+)}}^2}}\left(Y^2\left(k_T^2-2e_2+Y^2-3X_1X_2\right)+3(k_1k_2+X_1)(k_3k_4+X_2)\right)\right.\\&\nonumber\left.+\frac{k_T+2X_2^{(+)}}{4Y{X_2^{(+)}}^2}\left(X_1^{(+)}(X_1X_1^{(+)}+Y^2-k_1k_2)-Yk_1k_2\right)\left(X_2^{(-)}(X_2X_2^{(-)}+Y^2-k_3k_4)+Yk_3k_4\right)\right.\\&\nonumber\left.+\frac{k_T+2X_1^{(+)}}{4Y{X_1^{(+)}}^2}\left(X_1^{(-)}(X_1X_1^{(-)}+Y^2-k_1k_2)+Yk_1k_2\right)\left(X_2^{(+)}(X_2X_2^{(+)}+Y^2-k_3k_4)-Yk_3k_4\right)\right].
\end{align}
It is once again possible to confirm that this agrees with the result derived by the explicit evaluation of the Feynman integrals. However, the contributions from the two time orderings are individually divergent and so taking the limit inside this integral is quite involved. The weight-shifting approach is therefore more straightforward than evaluating this integral.

%%%%%%%%%%%%%%%%%%%%%%%%%%%%%%%%%%%%%%%%%%%%%%%%%%%%%%%%%%%%%%%

\section{Conclusions} 
Weight-shifting operators have proven to be efficient tools in various bootstrap programs. Both in the cosmological and in the conformal bootstrap, they have been used to construct correlators with altered spin or scaling dimension from simpler seed correlators~\cite{Costa:2011mg,Costa:2011dw,Costa:2016hju, Simmons-Duffin:2012juh,Karateev:2017jgd,Iliesiu:2015qra,Baumann:2019oyu}. In this paper, we have used a set of master integrals first defined in \cite{massivekinflow} to construct weight-shifting matrices that implement integer shifts in weight for cosmological correlators.

\vskip4pt

There are several benefits to this approach. First, to compute wavefunction coefficients with shifted weights, our procedure employs simple matrix multiplication which, especially for low-dimensional cases, is operationally simpler than the action of differential operators. This is particularly pronounced here as the mode functions of massive fields are special functions. Moreover, the structure in the master integrals allows us to define these operators on individual edges of a given Feynman diagram and therefore straightforwardly extend them to arbitrary diagrams. In principle, such a generalisation is possible for the previous derivative operators, but no procedure for generating these operators has previously been presented in the literature.
For example, the operators in~\cite{Arkani-Hamed:2018kmz} are defined to act on pairs of external legs, not individual ones. This makes computing diagrams with odd numbers of shifted edges a surprisingly challenging problem, a problem which does not arise in our methodology. 
Finally, since our construction is based on the master integrals, interactions involving derivatives can also be captured through matrix-implemented shifts. Consequently, generating such couplings does not require constructing additional weight-shifting operators.
\vskip4pt

It is worth noting that our construction applies equally well to momentum-space Witten diagrams in EAdS space. This follows from the fact that such Witten diagrams have the same structure as the connected part of the wavefunction coefficients 
considered here, with the Hankel mode functions replaced by modified Bessel functions. Since these mode functions satisfy the same differential equations as those in \reef{eqn: h differential equations} and obey analogous weight-shifting relations to those in \reef{eqn: h shift relations}, both the kinematic flow from \cite{massivekinflow} and the weight-shifting matrices presented here generalise straightforwardly to EAdS.
In this paper, we focus on de Sitter space because momentum-space correlators have received particular attention in the cosmology literature. Nevertheless, the weight-shifting procedure is, in some respects, even more powerful in EAdS than in de Sitter space. In particular, in EAdS, there is no need to contend with fields of imaginary weight, since unitarity requires real scaling dimensions. 

\vskip4pt

There are many avenues for further exploration of these weight-shifting matrices. The first, as hinted at in the introduction, is to combine the solutions to differential equations for  small deviations of conformally coupled $\xi$ with weight-shifting matrices. As explained in \cite{massivekinflow}, one can streamline the solutions to master integrals by iteratively solving order by order in small $\xi$. In addition, the weight-shifting matrices can be applied recursively to shift the weight to the closest integer. 
This results in a systematic and computationally efficient method to compute contributions to cosmological correlators for arbitrary masses. 
This approach, however, is limited to fields in the complementary series whose scaling dimensions are a real-valued shift from conformally coupled. It would be interesting to see if one could construct similar weight-shifting matrices that shift in the imaginary direction. 

\vskip4pt

Second, one of the main applications of weight-shifting operators in prior work is to produce spinning correlators from seed scalar correlators. In particular, in~\cite{Baumann:2019oyu} spin-raising and lowering operators are derived alongside weight-shifting operators. In principle, an equivalent procedure to what we have presented here should be applicable to such operators leading to spin-raising or lowering matrices. This is likely where the true power of these operators lies as, so far, kinematic flow has been restricted to interactions between scalar fields. Understanding such spin-raising operators would thus allow us to access bosonic fields of arbitrary spin in a similarly systematic manner.

\vskip3.5pt

Finally, it would be interesting to see if one can identify analogous weight-shifting matrices for Feynman diagrams involving loop integrals. Whilst our construction works naturally at the level of the loop integrand, the structure of such weight-shifting matrices for the integrated loop diagrams remains an open question.

%\vspace{-0.1cm}
 \paragraph{Acknowledgments} We would like to thank Daniel Baumann, Priyesh Chakraborty, Calvin Chen, Carolina Figueiredo, Hayden Lee and Saiei-Jaeyeong Matsubara-Heo for helpful discussions. 
We would  also like to thank Daniel Baumann for comments on the draft.
 \vskip 4pt
 \noindent
 KSV is supported  by the Dutch Research Council (NWO  \raisebox{-3pt}{\includegraphics[height=0.9\baselineskip]{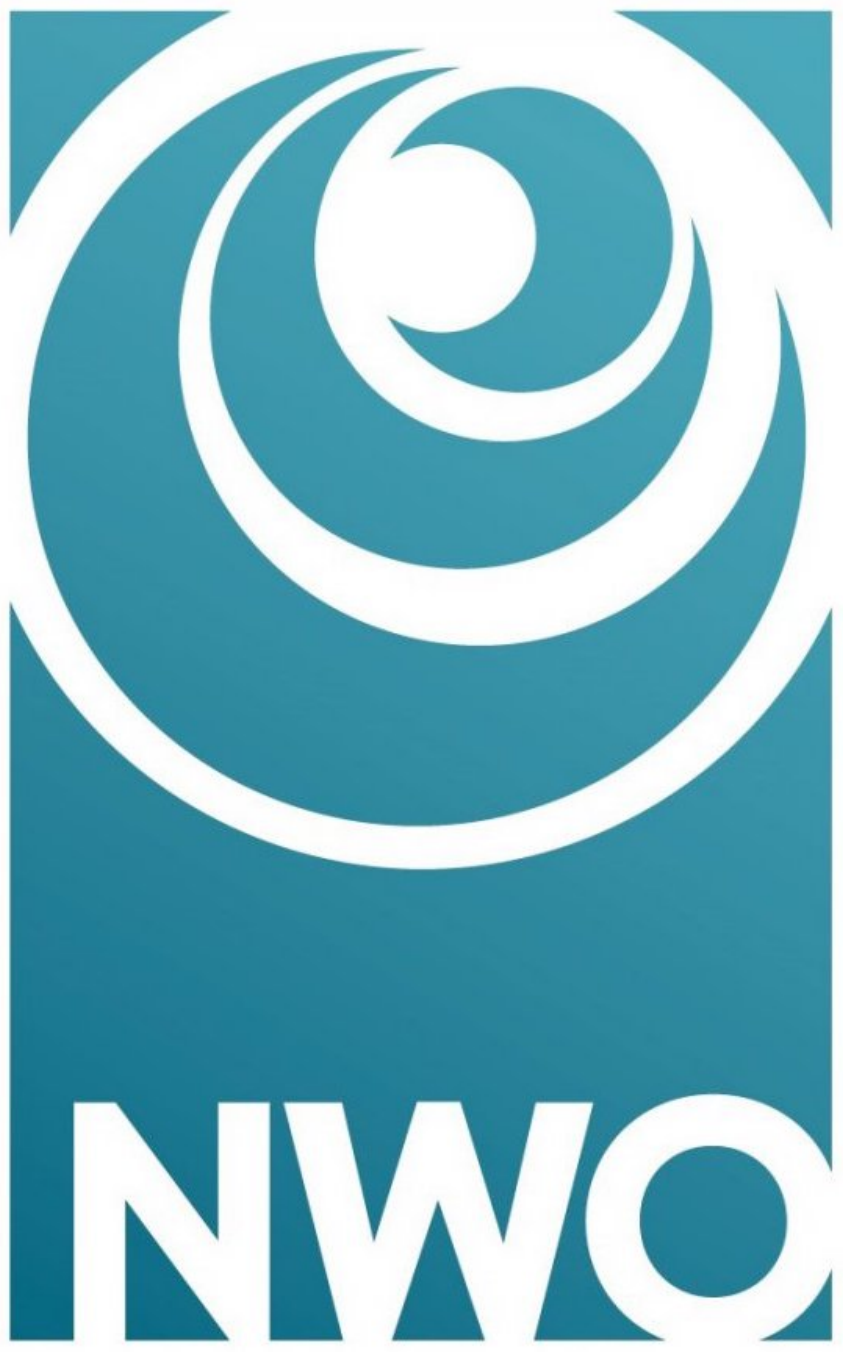}}) under {\hypersetup{urlcolor=black}\href {https://doi.org/10.61686/PRTOQ68396}{the grant}} for the project \textit{Constraining the Space of Cosmological Theories} with file number VI.Veni.242.438 of the research programme NWO Talent Programme, Veni.
The work of CdK and NW is funded by the European Union (ERC, UNIVERSE PLUS, 101118787). Views and opinions expressed are, however, those of the author(s) only and do not necessarily reflect those of the European Union or the European Research Council Executive Agency. Neither the European Union nor the granting authority can be held responsible for them.
HG is supported by a Postdoctoral Fellowship at National Taiwan University funded by the National Science and Technology Council (NSTC) 114-2923-M-002-011-MY5 Max Planck-IAS-NTU Center for Particle Physics, Cosmology and Geometry.

\newpage
\appendix
%%%%%%%%%%%%%%%%%%%%%%%%%%%%%%%%%%%%%%%%%%%%%%%%%%%%%%%%%%%%%%%

\section{Hyperplane Arrangement Representation}\label{app:Hyperplane}
In the main body of the text, we constructed weight-shifting matrices using the shift relations \eqref{eqn: h shift relations} satisfied by the $h_\nu^\pm$ and $\bar{h}^\pm_\nu$, as well as the $A$-matrices from \cite{massivekinflow}. Using the integral representations given below, we can construct the same weight-shifting matrices through carefully chosen integration-by-parts (IBP) and partial fraction decompositions (PFDs) alone. We show how this is done explicitly in the case of a contact diagram. The generalisation to arbitrary tree-level diagrams is straightforward, though tedious.

\vskip4pt

As described in \cite[§3, Appendix B]{massivekinflow}  we have the following integral representations
\ba
    h^\pm_\nu(k,\eta)&=\frac{-2^\xi}{\Gamma[-\xi]k^\xi}\int_{\gamma_1} \ud t(t^2-1)^{-\xi}\frac{e^{ik\eta t}}{1\mp t}\,,\\
    \bar{h}_\nu^\pm(k,\eta)&=\frac{-2^\xi}{\Gamma[-\xi]k^\xi}\int_{\gamma_1} \ud t(t^2-1)^{-\xi}\frac{e^{-ik\eta t}}{1\pm t}\,,\\
    (-\eta)^{\alpha}&=\frac{e^{-i\frac{\pi}{2}\alpha}}{\Gamma(-\alpha)}\int_{\gamma_0} \ud\omega \omega^{-\alpha-1} e^{i\eta \omega}\,.
\ea
In practice, to ensure that these integral representations are well-defined for the values of $\alpha$ and $\nu$ that we are interested in, we will work with analytic continuations of them. These analytic continuations have boundary-less integration contours, $\gamma_x$, which avoids the branch-points at $x$ and approaches $\infty$ from a direction that ensures the vanishing of the integral through an appropriate $i\epsilon$ prescription. We will use this fact frequently below, to justify setting boundary terms to zero. We refer the reader to \cite{eulerintegrals} for further details on this procedure.

\vskip4pt

\paragraph{Contact diagram} Using the integral representations above, we first write the relevant master integrals as twisted integrals on hyperplane arrangements
\begin{equation}
    \begin{split}
        \psi^{\pm}_{\alpha,\nu}&=\int_{-\infty}^0\frac{\ud\eta}{(-\eta)^{1+\alpha+\xi}}e^{iX\eta}h_\nu^\pm(k,\eta)\\
        &=N_{\alpha,\nu}\int_{\gamma_1}\ud t\int_{\gamma_0}\ud\omega\frac{\omega^{\alpha+\xi}(t^2-1)^{-\xi}}{(X+\omega+kt)(1\mp t)}\,, \quad N_{\alpha,\nu}:=\frac{i2^\xi e^{i\frac{\pi}{2}(1+\alpha+\xi)}}{\Gamma[-\xi]\Gamma[1+\alpha+\xi]k^{\xi}}\,.
    \end{split}
\end{equation}
In what follows, we will not write the integral contour explicitly. The aim is to derive the matrices $M$ and $\mathcal{M}$ that implement the shifts
\begin{equation}
    \vec I_{\alpha,\nu+1}= {M} \cdot \vec I_{\alpha,\nu}\, \quad\text{and}\quad \vec I_{\alpha,\nu+1}= \mathcal{M} \cdot \vec I_{\alpha+1,\nu}\,,
\end{equation}
where, as in the main text, we have defined the vector $\vec{I}_{\alpha,\nu}=[\psi_{\alpha,\nu}^+\quad\psi^-_{\alpha,\nu}]^T$. To do this, we employ the following three relations, each derived using a combination of IBP and PFDs.

\vskip8pt

\noindent\textit{Shifting $\alpha$ using derivatives:}\newline
\noindent Using Stokes' theorem, we have that
\begin{equation}
    0=N_{\alpha,\nu}\int\ud t\ud\omega\frac{\partial}{\partial\omega}\left[\frac{\omega^{\alpha+\xi}(t^2-1)^{-\xi}}{(X+\omega+kt)(1\mp t)}\right]\,.
\end{equation}
Taking the derivative and using that $N_{\alpha-1,\nu}=-i(\alpha+\xi)N_{\alpha,\nu}$, we find that the right-hand side of the above is equal to
\begin{equation}
   N_{\alpha,\nu}\int\ud t\ud\omega\frac{\omega^{\alpha+\xi}(t^2-1)^{-\xi}}{(X+\omega+kt)(1\mp t)}\left[\frac{\alpha+\xi}{\omega}-\frac{1}{X+\omega+kt}\right]=i\psi^{\pm}_{\alpha-1,\nu}+\partial_X\psi^\pm_{\alpha,\nu}\,,
\end{equation}
and so 
\begin{equation}\label{eqn: alpha shift through derivative}
-i\psi^{\pm}_{\alpha-1,\nu}=\partial_X\psi^\pm_{\alpha,\nu}\,.  
\end{equation}

\noindent\textit{Shifting $\alpha$ by linear combination:}\newline
Using that $N_{\alpha-1,\nu}=-i(\alpha+\xi)N_{\alpha,\nu}$, we have the following integral representation for the functions $\psi^\pm_{\alpha-1,\nu}$,
\begin{equation}
    \psi^\pm_{\alpha-1,\nu}=-i(\alpha+\xi)N_{\alpha,\nu}\int\ud t\ud\omega\frac{\omega^{\alpha+\xi-1}(t^2-1)^{-\xi}}{(X+\omega+kt)(1\mp t)}\,.
\end{equation}
Using the PFD
\begin{equation}
    \frac{1}{\omega(X+\omega+kt)(1\mp t)}=\frac{1}{X\pm k}\left[\frac{1}{\omega(1\mp t)}-\frac{1}{(X+\omega+kt)(1\mp t)}\pm\frac{k}{\omega(X+\omega+kt)}\right]\,,
\end{equation}
We can write this as
\begin{equation}\label{eqn: alpha shift PFD}
    \psi^\pm_{\alpha-1,\nu}=-i\frac{\alpha+\xi}{X\pm k}N_{\alpha,\nu}\int\ud t\ud\omega\frac{\omega^{\alpha+\xi}}{(t^2-1)^{\xi}}\left[\frac{1}{\omega(1\mp t)}-\frac{1}{(X+\omega+kt)(1\mp t)}\pm\frac{k}{\omega(X+\omega+kt)}\right]\,.
\end{equation}
The first term can we re-cast as a total derivative term in $\omega$ and is set to zero via Stokes' theorem. The second term is proportional to $\psi^\pm_{\alpha,\nu}$. The third term requires further manipulation. Focusing on the third term, note that it can be re-written in the form
\begin{equation}
   \frac{\mp i}{X\pm k}N_{\alpha,\nu}\int\ud t\ud\omega(t^2-1)^{-\xi}\frac{k}{(X+\omega+kt)}\frac{\partial}{\partial\omega}(\omega^{\alpha+\xi})\,.
\end{equation}
Using IBP, this is equal to
\begin{equation}
   \frac{\pm i}{X\pm k}N_{\alpha,\nu}\int\ud t\ud\omega\omega^{\alpha+\xi}(t^2-1)^{-\xi}\frac{\partial}{\partial\omega}\left[\frac{k}{(X+\omega+kt)}\right]\,.
\end{equation}
The result of taking this derivative with respect to $\omega$ is, up to a factor of $k$, identical to what we would get if instead we took a derivative with respect to $t$. This means we can write the above as 
\begin{equation}
   \frac{\pm i}{X\pm k}N_{\alpha,\nu}\int\ud t\ud\omega\omega^{\alpha+\xi}(t^2-1)^{-\xi}\frac{\partial}{\partial t}\left[\frac{1}{(X+\omega+kt)}\right]\,.
\end{equation}
Again employing IBP, this time in $t$, we find that the above is equal to
\begin{equation}
\begin{split}
    &\frac{\pm i\xi}{X\pm k}N_{\alpha,\nu}\int\ud t\ud\omega\frac{\omega^{\alpha+\xi}(t^2-1)^{-\xi}}{(X+\omega+kt)}\frac{2t}{t^2-1}\\
    &\quad = \frac{\pm i\xi}{X\pm k}N_{\alpha,\nu}\int\ud t\ud\omega\frac{\omega^{\alpha+\xi}(t^2-1)^{-\xi}}{(X+\omega+kt)}\left[\frac{1}{1+t}-\frac{1}{1-t}\right]\,.
\end{split}
\end{equation}
The first term in the second line is proportional to $\psi^+_{\alpha,\nu}$ whilst the second term is proportional to $\psi^-_{\alpha,\nu}$. Substituting everything back into \eqref{eqn: alpha shift PFD}, we find the following shift relation
\begin{equation}\label{eqn: alpha lin comb shift}
\begin{split}
    \psi^\pm_{\alpha-1,\nu}&=i\frac{\alpha+\xi}{X\pm k}\psi^\pm_{\alpha,\nu}\pm \frac{i\xi}{X\pm k}\left[\psi^-_{\alpha,\nu}-\psi^+_{\alpha,\nu}\right]\\
    &=\frac{i}{X\pm k}[\alpha\psi^\pm_{\alpha,\nu}+\xi\psi^\mp_{\alpha,\nu}]\,.
\end{split}
\end{equation}
In particular, this shift relation allows us to replace the partial derivative with respect to $X$ in \eqref{eqn: alpha shift through derivative} with a linear combination of $\psi^+_{\alpha,\nu}$ and $\psi^-_{\alpha,\nu}$. Note that this is just an alternative approach to constructing shift relation one arrives at using the $A$-matrices of~\cite{massivekinflow} introduced in Section~\ref{subsubsec: A-matrix alpha shift}. Note that by taking appropriate linear combinations of the expression above, we arrive at
\begin{equation}\label{eqn: alpha lin comb shift inverse}
    (\alpha^2-\xi^2)\psi^\pm_{\alpha,\nu}=-i\alpha(X\pm k)\psi^\pm_{\alpha-1,\nu}+i\xi(X\mp k)\psi^\mp_{\alpha-1,\nu}\,,
\end{equation}
which in some sense gives the inverse of \eqref{eqn: alpha lin comb shift}.

\vskip8pt

\noindent\textit{Shifting $\nu$:}\newline
\noindent By Stokes' theorem, we have that
\begin{equation}
    0=N_{\alpha,\nu}\int\ud t\ud\omega\frac{\partial}{\partial t}\left[\frac{\omega^{\alpha+\xi}(t^2-1)^{-\xi}}{(X+\omega+kt)(1\mp t)}\right]\,.
\end{equation}
Carrying out the differentiation, we find that the right hand side of the above can be written in the form
\begin{equation}
    N_{\alpha,\nu}\int\ud t\ud\omega\frac{\omega^{\alpha+\xi}(t^2-1)^{-\xi}}{(X+\omega+kt)}\left[-\frac{2\xi t}{(t^2-1)(1\mp t)}-\frac{k}{(X+\omega+kt)(1\mp t)}\pm\frac{1}{(1\mp t)^2}\right]\,.
\end{equation}
Using an alternative PFD, we can write the terms in the brackets as follows
\begin{equation}
    \mp\frac{(1+2\xi)}{2(1\pm t)}\mp\frac{(1+2\xi)}{2(1\mp t)}\mp\frac{2(\xi+1)}{(t^2-1)(1\mp t)}-\frac{k}{(X+\omega+kt)(1\mp t)}\,.
\end{equation}
Substituting this back into the integral, using that $N_{\alpha-1,\nu+1}=-2(\xi+1)N_{\alpha,\nu}/k$ and recalling that we define $\xi=\nu-1/2$, we find
\begin{equation}
    0=\nu\psi^{\mp}_{\alpha,\nu}+\nu\psi^{\pm}_{\alpha,\nu}\mp k\partial_X\psi^\pm_{\alpha,\nu}-k\psi^{\pm}_{\alpha-1,\nu+1}\,.
\end{equation}
Using \eqref{eqn: alpha shift through derivative}, we can replace the derivative with respect to $X$ with a shifted $\alpha$ term to get
\begin{equation}\label{eqn: unprocessed nu shift}
    0=\nu\psi^{\mp}_{\alpha,\nu}+\nu\psi^{\pm}_{\alpha,\nu}\pm i k\psi^\pm_{\alpha-1,\nu}-k\psi^{\pm}_{\alpha-1,\nu+1}\,.
\end{equation}

\noindent\textit{Constructing $\mathcal{M}$:}\newline
\noindent Using \eqref{eqn: alpha lin comb shift}, we can replace the $\psi^\pm_{\alpha-1,\nu}$ in \eqref{eqn: unprocessed nu shift} with a linear combination of $\psi^\pm_{\alpha,\nu}$. This yields:
\begin{equation}
    0=\nu\psi^{\mp}_{\alpha,\nu}+\nu\psi^{\pm}_{\alpha,\nu}\mp\frac{k}{X\mp k}\left[\alpha\psi^\pm_{\alpha,\nu}+\xi\psi^\mp_{\alpha,\nu}\right] -k\psi^{\pm}_{\alpha-1,\nu+1}\,.
\end{equation}
Replacing each $\alpha$ by $\alpha+1$ and rearranging, this is
\ba
    \psi^{+}_{\alpha,\nu+1}&=\left[\frac{\nu}{k}-\frac{\alpha+1}{X+k}\right]\psi^+_{\alpha+1,\nu}+\left[\frac{\nu}{k}-\frac{\xi}{X+k}\right]\psi^-_{\alpha+1,\nu}\,,\\
    \psi^{-}_{\alpha,\nu+1}&=\left[\frac{\nu}{k}+\frac{\xi}{X-k}\right]\psi^+_{\alpha+1,\nu}+\left[\frac{\nu}{k}+\frac{\alpha+1}{X-k}\right]\psi^-_{\alpha+1,\nu}\,,
\ea
This can be arranged as the following matrix equation
\begin{equation}
    \vec{I}_{\alpha,\nu+1}=\mathcal{M}\cdot\vec{I}_{\alpha+1,\nu}\,,\quad\text{where}\quad\mathcal{M}=\begin{bmatrix}\frac{\nu}{k}-\frac{\alpha+1}{X+k}&\frac{\nu}{k}-\frac{\xi}{X+k}\\
    \frac{\nu}{k}+\frac{\xi}{X-k}&\frac{\nu}{k}+\frac{\alpha+1}{X-k}  
    \end{bmatrix}.
\end{equation}
Note that this $\mathcal{M}$ is identical to the matrix on the right in \eqref{eq: M for contact}.

\vskip8pt

\noindent\textit{Constructing $M$:}\newline
\noindent This time, we use \eqref{eqn: alpha lin comb shift inverse} to replace the $\psi^\pm_{\alpha,\nu}$ in \eqref{eqn: unprocessed nu shift} with linear combinations of $\psi^\pm_{\alpha-1,\nu}$. We find
\begin{equation}
    0=\frac{i\nu}{\alpha+\xi}\left[\psi^\mp_{\alpha-1,\nu}(X\mp k)+\psi^\pm_{\alpha-1,\nu}(X\pm k)\right]\mp ik\psi^\pm_{\alpha-1,\nu}+k\psi^\pm_{\alpha-1,\nu+1}\,.
\end{equation}
Replacing $\alpha$ by $\alpha+1$ and rearranging, this becomes
\begin{equation}
    \psi^\pm_{\alpha,\nu+1}=-i\frac{\nu}{k}\frac{1}{1+\alpha+\xi}\left[\psi^\pm_{\alpha,\nu}(X\pm k)+\psi^\mp_{\alpha,\nu}(X\mp k)\right]\pm i\psi^\pm_{\alpha,\nu}\,.
\end{equation}
This can be arranged as the matrix equation
\begin{equation}
    \vec{I}_{\alpha,\nu+1}=M\cdot\vec{I}_{\alpha,\nu}\,,\quad\text{where}\quad iM=\begin{bmatrix}\frac{\nu}{k}\frac{X+k}{1+\alpha+\xi}-1&\frac{\nu}{k}\frac{X-k}{1+\alpha+\xi}\\\frac{\nu}{k}\frac{X+k}{1+\alpha+\xi}&\frac{\nu}{k}\frac{X-k}{1+\alpha+\xi}+1 
    \end{bmatrix}.
\end{equation}
Note that this $M$ is identical to the matrix on the left in \eqref{eq: M for contact}.

\section{Weight-Shifting Differential Operator}\label{app:derivativews}

In this appendix, we review the approach taken in~\cite{Arkani-Hamed:2018kmz} to finding the weight-shifting operators from the second order differential equations satisfied by the single exchange diagram. We will then discuss how this approach can be applied to the linear differential equation system and demonstrate that this results in the same operator as we found in the main text.

\vskip4pt

The four-point exchange diagram in~\eqref{eq: exchange} with conformally coupled external edges satisfies a second order differential equation
\ba
    \left[\Delta_U+\frac{(1+2\epsilon)^2}{4}-\nu^2\right]\psi_{\epsilon,\epsilon,\nu}(U,V)=C_0\left(\frac{1}{U+V}\right)^{1+2\epsilon}\,,\\ \Delta_U=(U^2-1)\partial_U^2+2U(1+\epsilon)\partial_U\,.
\ea
where $U=\frac{X_1}{Y},\, V=\frac{X_2}{Y}$, $d=3+2\epsilon$ and we can find an equivalent expression for derivatives with respect to $V$. The weight-shifting operator in this case is
\ba
    M_{\epsilon\epsilon,\nu} : \psi_{\epsilon,\epsilon,\nu}\rightarrow\psi_{\epsilon,\epsilon,\nu+1}\,.
\ea
Before we construct this full weight-shifting operator, we first consider the contact example which satisfies a homogeneous differential equation
\ba
    \left[\Delta_U+\frac{(1+2\epsilon)^2}{4}-\nu^2\right]\psi_{\epsilon,\nu}(U)=0\,.
\ea
The weight-shifting operator in this case can be derived by considering its action on this differential equation
\ba
    M_{\epsilon,\nu}\left[\Delta_U+\frac{(1+2\epsilon)^2}{4}-\nu^2\right]\psi_{\epsilon,\nu}(U)&=0\,.
\ea
By enforcing that $M_{\epsilon,\nu}\psi_{\epsilon,\nu}(U)$ is a solution to this equation for a shifted $\nu$, we find that
\ba\label{eq:ShiftDE}
    \left[M_{\epsilon,\nu},\Delta_u\right]\psi_{\epsilon,\nu}(U)=-(2\nu+1)M_{\epsilon,\nu}\psi_{\epsilon,\nu}(U)\,.
\ea
Introducing an ansatz of the form
\ba
    M_{\epsilon,\nu}=f(U)\partial_U+g(U)\,,
\ea
gives us a pair of coupled second order differential equations for $f$ and $g$ which we solve assuming a power series solution and fixing the coefficients through a recursion relation. There are four possible solutions depending on the choice of the first few coefficients in this power series. However, there is just one solution that gives us a finite polynomial for both $f$ and $g$ which is
\ba
    f(U)&=f_0(1-U^2)\,,& g(U)&=-\frac{1}{2}(2\epsilon+1+2\nu)f_0\,.
\ea
We are left with the freedom to specify some overall constant $f_0$. This reflects the fact that rescaling the solutions to a differential equation by a constant just gives a new solution to the equation. For the exchange diagram, we consider the operator
\ba
    M_{\epsilon,\epsilon,\nu}(U,V)\psi_{\epsilon,\epsilon,\nu+1} =M_{\epsilon,\nu}(U)M_{\epsilon,\nu}(V)\psi_{\epsilon,\epsilon,\nu}+F(U,V)\,.
\ea
In principle, one could construct and solve a differential equation for $F$ from an equivalent expression to~\eqref{eq:ShiftDE}, but it turns out to be more efficient to extract this directly from the integral,
\ba
   M_{\epsilon,\nu}(U)\psi_{\epsilon,\epsilon,\nu} &= \frac{i f_0}{2}\lambda^2 Y^{2\alpha}\int \ud x_1 \ud x_2  \tilde{G}_\nu(x_1,x_2) (2x_1(1+\partial_{x_1}^2)+2(2+\epsilon+\nu)\partial_{x_1})e^{-iUx_1}e^{-iVx_2}\\&=\frac{i f_0}{2}\lambda^2 Y^{2\alpha}\int \ud x_1 \ud x_2  e^{-iUx_1}e^{-iVx_2}(2x_1-(3+2\nu+2\epsilon)\partial_{x_1}+2x_1\partial_{x_1}^2)\tilde{G}_\nu(x_1,x_2) \,,
\ea
where we have combined the factors of $x_i^\epsilon$ into a new Green's function $\tilde{G}_\nu$, which satisfies the differential equation
\ba
    ix_1^{d-3}\delta(x_1-x_2)&=x_1^2\partial_{x_1}^2 \tilde{G}_\nu+x_1(1-2\alpha)\partial_{x_1}\tilde{G}_\nu+(x_1^2-\nu^2+\alpha^2) \tilde{G}_\nu\\
    \tilde{G}_\nu(x_1,x_2)&=\frac{\pi}{4(x_1x_2)^{1/2-\epsilon}}\left(H_{\nu}^{(1)}(x_1)H_\nu^{(2)}(x_2)\theta_{21}+H_\nu^{(1)}(x_2)H_{\nu}^{(2)}(x_1)\theta_{12}\right).
\ea
We use this to eliminate the second derivative of $\tilde{G}$, in exchange for the introduction of a $\delta$-function that collapses one of the integrals
\ba
   M_{\epsilon,\nu}(U)&\psi_{\epsilon,\epsilon,\nu} = \frac{i f_0}{2}\lambda^2 Y^{2\alpha}\left[-i\int \frac{\ud x_1}{x_1^{1-2\epsilon}} e^{-i(U+V)x_1} \right.\\&\left.+\int \ud x_1 \ud x_2 \frac{2\epsilon-2\nu-1}{4x_1}e^{-i(Ux_1+Vx_2)}\left( (2\epsilon+2\nu-1)-2x_1\partial_{x_1}\right)\tilde{G}_\nu(x_1,x_2)\right].
\ea
The derivative operator in this second line shifts $\nu\rightarrow \nu+1$ on the first vertex. Applying the second weight-shifting operator completes the shift on $\tilde{G}_\nu$ and also modifies the boundary term,
\ba
   M_{\epsilon,\nu}(V)M_{\epsilon,\nu}(U)\psi_{\epsilon,\epsilon,\nu} &= f_0^2\lambda^2 Y^{2\alpha}\left[\int \frac{\ud x_1}{x_1^{1-2\epsilon}} e^{-i(U+V)x_1} (i(U^2-1)x_1-2U\epsilon)\right.\\&\left.-\int \ud x_1 \ud x_2 \frac{(2\epsilon-2\nu-1)^2}{4}e^{-i(Ux_1+Vx_2)}\tilde{G}_{\nu+1}(x_1,x_2)\right].
\ea
Evaluating this integral then gives us an expression for $F(U,V)$. Furthermore, we can compare the integral in the final line to $\psi_{\epsilon,\epsilon,\nu+1}$, which also fixes $f_0$, 
\ba
    F(U,V)=-f_0^2C_0\left(\frac{1}{U+V}\right)^{1+2\epsilon}\left(1+UV\right)\,,\qquad f_0=\frac{2i}{2\epsilon-2\nu-1}\,.
\ea
This exactly matches the weight-shifting operator found in~\cite{Arkani-Hamed:2018kmz} for $d=3$. 

\vskip4pt

This operator cannot be used directly on our master integrals as they satisfy different differential equations,
\ba
    \Delta^\pm_U \psi^{\pm\pm}_{\epsilon,\epsilon,\nu}&=\mp(2\nu-1) J_{\epsilon,\epsilon}\,,\\
    \Delta^\pm_U \psi^{\pm\mp}_{(\epsilon,\epsilon),\nu}&=\frac{(2\pm(U-V))(2\nu-1+2\epsilon)}{U+V} I_{2\epsilon}\,,
\ea
for the sake of compactness we have introduced a new differential operator
\ba
    \Delta^\pm_U=(U^2-1)\partial_U^2+(U\mp 1-2U\epsilon)\partial_U+\left(\epsilon^2-\left(\nu-\frac{1}{2}\right)^2\right)\,.
\ea
Following the same steps as before, we find the homogeneous weight-shifting operator as
\ba
    M^\pm_{\epsilon,\nu}(U)=A\left[\left(\mp\frac{1}{2\nu}+U\right)(\nu-\frac{1}{2}+\epsilon)+(U^2-1)\partial_U\right],\qquad A=\frac{4\nu}{i(2\nu-1)(2\nu+1+2\epsilon)}\,.
\ea
Using the kinematic flow, we can recast these derivative weight-shifting operators for our master integrals into matrices,
\ba
    iM_{\epsilon,\nu}\vec{I}_{\epsilon,\nu}&=\left(\frac{\left(2\nu U\mp1\right)(\xi+\epsilon)}{2\xi(\xi+1+\epsilon)}\begin{bmatrix}
        1&0\\0&1
    \end{bmatrix}+\frac{2\xi+1 }{2\xi(\xi+1+\epsilon)}\begin{bmatrix}
        -\epsilon(U-1)&\xi(U-1)\\\xi(U+1)&-\epsilon(U+1)
    \end{bmatrix}\right)\cdot\vec{I}_{\epsilon,\nu}\,.
\ea
After some regrouping of terms this recovers exactly the results found in the main text in~\eqref{eq: M for contact}.

\vskip4pt

To obtain the exchange weight-shifting operator in this derivative form, we fix the boundary term by acting with the homogeneous weight-shifting operator twice,
\ba
    M^\pm_{\epsilon,\nu}(U)M^\pm_{\epsilon,\nu}(V)\psi^{\pm\pm}_{\epsilon,\epsilon,\nu}&=\psi^{\pm\pm}_{\epsilon,\epsilon,\nu+1}\pm A^2(1+2\nu)(1+UV\mp\frac{U+V}{2\nu})J_{\epsilon,\epsilon}\,,\\M^\pm_{\epsilon,\nu}(U)M^\mp_{\epsilon,\nu}(V)\psi^{\pm\mp}_{\epsilon,\epsilon,\nu}&=\psi^{\pm\mp}_{\epsilon,\epsilon,\nu+1}+\frac{ A^2\epsilon((U-V)^2-4-2(1+UV)(1\pm\nu(U-V)))}{\nu(U+V)}J_{\epsilon,\epsilon}\,.
\ea
This can, just as in the homogeneous case be rewritten as a matrix using the first order system of differential equations to replace all the derivatives with respect to the kinematic variables. This exactly recovers the weight-shifting operators presented in the main text of this paper. We highlight this equivalence to make it clear that our matrices do not just achieve the same goal as the operators in~\cite{Arkani-Hamed:2018kmz} but are, in-fact, the same operators applied directly at the level of the master integrals.

%%%%%%%%%%%%%%%%%%%%%%%%%%%%%%%%%%%%%%
\section{Shift Relations using Kronecker Products}\label{app: n=2 contact nu shift example}
In the main text, we presented the expression \eqref{eqn: generic shifting 1} as a generalisation of the relation \eqref{eq: intermediate I}. In this appendix, we show this explicitly in the case of the $n=2$ contact diagram, thereby demonstrating the power of the Kronecker product as an organisational tool for structuring calculations. The master integrals for the $n=2$ contact diagram can be organised into the following $4$-entry vector
\begin{equation}
    \vec{I}_{\alpha,\bm{\nu}}=\int\frac{\ud\eta}{(-\eta)^{1+\alpha+\xi_1+\xi_2}}e^{iX\eta}\begin{bmatrix}h_{\nu_1}^+(k_1,\eta)\\h_{\nu_1}^-(k_1,\eta)\end{bmatrix}\otimes\begin{bmatrix}h_{\nu_2}^+(k_2,\eta)\\h_{\nu_2}^-(k_2,\eta)\end{bmatrix}.
\end{equation}
Now suppose we want to realise the shift $\nu_2\rightarrow\nu_2+1$ in $\vec{I}_{\alpha,\bm{\nu}}$ whilst keeping $\nu_1$ constant. Multiplying $\vec{I}_{\alpha,\bm{\nu}}$ by $iB_2$ gives
\begin{equation}
\begin{split}
    iB_2\vec{I}_{\alpha,\bm{\nu}}&=\int\frac{\ud \eta}{(-\eta)^{1+\alpha+\xi_1+\xi_2}}e^{iX\eta}\left(\mathbbm{I}_2\otimes iB\right)\begin{bmatrix}h_{\nu_1}^+(k_1,\eta)\\h_{\nu_1}^-(k_1,\eta)\end{bmatrix}\otimes\begin{bmatrix}h_{\nu_2}^+(k_2,\eta)\\h_{\nu_2}^-(k_2,\eta)\end{bmatrix}\\
    &=\int\frac{\ud\eta}{(-\eta)^{1+\alpha+\xi_1+\xi_2}}e^{iX\eta}\begin{bmatrix}h_{\nu_1}^+(k_1,\eta)\\h_{\nu_1}^-(k_1,\eta)\end{bmatrix}\otimes\left(iB\cdot\begin{bmatrix}h_{\nu_2}^+(k_2,\eta)\\h_{\nu_2}^-(k_2,\eta)\end{bmatrix}\right),
    \end{split}
\end{equation}
where in the second line we have used the mixed product property of the Kronecker product. Similarly, multiplying $\vec{I}_{\alpha+1,\bm{\nu}}$ by $\nu_2C_2/k_2$ gives
\begin{equation}
\begin{split}
    \frac{\nu_2}{k_2}C_2\vec{I}_{\alpha+1,\bm{\nu}}&=\int\frac{\ud\eta}{(-\eta)^{2+\alpha+\xi_1+\xi_2}}e^{iX\eta}\left(\mathbbm{I}_2\otimes\frac{\nu_2}{k_2}C\right)\begin{bmatrix}h_{\nu_1}^+(k_1,\eta)\\h_{\nu_1}^-(k_1,\eta)\end{bmatrix}\otimes\begin{bmatrix}h_{\nu_2}^+(k_2,\eta)\\h_{\nu_2}^-(k_2,\eta)\end{bmatrix}\\
    &=\int\frac{\ud\eta}{(-\eta)^{1+\alpha+\xi_1+\xi_2}}e^{iX\eta}\begin{bmatrix}h_{\nu_1}^+(k_1,\eta)\\h_{\nu_1}^-(k_1,\eta)\end{bmatrix}\otimes\left(-\frac{\nu_2}{k_2\eta}C\cdot\begin{bmatrix}h_{\nu_2}^+(k_2,\eta)\\h_{\nu_2}^-(k_2,\eta)\end{bmatrix}\right),
\end{split}
\end{equation}
where, in addition, we have used that the Kronecker product is bilinear. Adding these two yields
\begin{equation}
\begin{split}
    iB_2\vec{I}_{\alpha,\bm{\nu}}+\frac{\nu_2}{k_2}C_2\vec{I}_{\alpha+1,\bm{\nu}}&=\int\frac{\ud\eta}{(-\eta)^{1+\alpha+\xi_1+\xi_2}}e^{iX\eta}\begin{bmatrix}h_{\nu_1}^+(k_1,\eta)\\h_{\nu_1}^-(k_1,\eta)\end{bmatrix}\otimes\left(\left[iB-\frac{\nu_2}{k_2\eta}C\right]\begin{bmatrix}h_{\nu_2}^+(k_2,\eta)\\h_{\nu_2}^-(k_2,\eta)\end{bmatrix}\right).
\end{split}
\end{equation}
 But then, using \reef{eqn: shift relations}, we find that
 \begin{equation}
 \begin{split}
     iB_2\vec{I}_{\alpha,\bm{\nu}}+\frac{\nu_2}{k_2}C_2\vec{I}_{\alpha+1,\bm{\nu}}&=\int\frac{\ud\eta}{(-\eta)^{1+\alpha+\xi_1+\xi_2}}e^{iX\eta}\begin{bmatrix}h_{\nu_1}^+(k_1,\eta)\\h_{\nu_1}^-(k_1,\eta)\end{bmatrix}\otimes\left(\frac{1}{(-\eta)}\begin{bmatrix}h_{\nu_2+1}^+(k_2,\eta)\\h_{\nu_2+1}^-(k_2,\eta)\end{bmatrix}\right)\\
     &=\int\frac{\ud\eta}{(-\eta)^{1+\alpha+\xi_1+(\xi_2+1)}}e^{iX\eta}\begin{bmatrix}h_{\nu_1}^+(k_1,\eta)\\h_{\nu_1}^-(k_1,\eta)\end{bmatrix}\otimes\begin{bmatrix}h_{\nu_2+1}^+(k_2,\eta)\\h_{\nu_2+1}^-(k_2,\eta)\end{bmatrix}\\
     &=\vec{I}_{\alpha,\bm{\nu}+\bm{e}_2}\,.
 \end{split}
 \end{equation}
This result exactly matches the expression in~\eqref{eq: intermediate I} except we have replaced the $B$ and $C$ matrices with similar matrices padded using our Kronecker product prescription.

\section{Closed Formula for Contact Diagrams}\label{app: derivation of contact weight-shifting}

In this section, we derive an explicit expression for the weight-shifting matrix $M$ corresponding to a generic contact diagram.
In the main text, we explained that the weight-shifting matrices $\mathcal{M}$ that shift both the relevant weight and vertex parameter(s) are easier to construct than the weight-shifting matrices $M$ that shift the weight alone. This is because the $\mathcal{M}$-matrices are products of $\partial_{X_i}$-components of the $A$-matrices introduced in \cite{massivekinflow} -- matrices which, for a given diagram, are constructed using a simple algorithm. In contrast, computing the $M$-matrices requires one to invert these $A$-matrix components. This quickly becomes intractable as one considers more and more complicated diagrams.

\vskip4pt

Rather than performing these matrix inversions for each diagram on a case-by-case basis, one may wonder if, just like the $A$-matrices themselves, the inverses of their $\partial_{X_i}$-components can similarly be constructed using some simple rule or formula that applies to arbitrary tree-level Feynman diagrams. As described in the main text, we believe that such a rule does exist, but that its intricacies make it unenlightening. We leave the details of the construction of such a general rule as an open problem for those interested. In this appendix, we expose some of the underlying structure in these matrix inverses by deriving a closed formula for the matrix $M^\text{ext}_j$. This matrix was first introduced in Section~\ref{subsec: generic shifting}, and realises the shift $\nu_j\rightarrow\nu_j+1$ on a contact diagram with $n$ arbitrary mass bulk-to-boundary propagators.

\vskip4pt

Since we know what $B_j$ is, computing $M_j^{\mathrm{ext}}$ amounts to computing $C_j[A^{(X)}_{\alpha+1}]^{-1}$. We do this by presenting an ansatz for $C_j[A^{(X)}_{\alpha+1}]^{-1}$, call this $N_j$, and then showing that $N_jA^{(X)}_{\alpha+1}=C_j$. This proof strategy relies on the invertibility of $A^{(X)}_{\alpha+1}$. If it were not invertible then there could be multiple matrices that, when multiplied with $A^{(X)}_{\alpha+1}$ on the left, give $C_j$. We will assume that values of the external kinematics, vertex parameter and weights are such that $A^{(X)}_{\alpha+1}$ is indeed invertible. Define the matrices
\begin{equation}
    F_{\bm{p}}=\bigotimes_{\ell=1}^n\begin{bmatrix}1&(-1)^{p_\ell}\\(-1)^{p_\ell}&1\end{bmatrix}, \ \ \ G=X\mathbbm{I}_{2^n}+\sum_{\ell=1}^nk_\ell B_\ell\,.
\end{equation}
Using these, we construct the ansatz
\begin{equation}
    N_j=\frac{1}{2^{n-1}}\sum_{\substack{\bm{p}\in\{0,1\}^n\\p_j=0}}\frac{1}{1+\alpha+\sum_{\ell=1}^n(-1)^{p_\ell}\xi_\ell}F_{\bm{p}}\cdot G\,.
\end{equation}
From the diagrammatic rules of \cite{massivekinflow}, we know that $A^{(X)}_{\alpha+1}$ takes the form
\begin{equation}
    A^{(X)}_{\alpha+1}=G^{-1}\cdot\bigg[(1+\alpha)\mathbbm{I}_{2^n}+\sum_{\ell=1}^n\xi_\ell H_\ell\bigg]\,,
\end{equation}
where we have defined
\begin{equation}
    H_\ell=\bigotimes_{s=1}^n\begin{bmatrix}1-\delta_{s\ell}&\delta_{s\ell}\\\delta_{s\ell}&1-\delta_{s\ell}\end{bmatrix}.
\end{equation}
We now derive the identities
\begin{align}
    F_{\bm{p}}H_\ell&=(-1)^{p_\ell}F_{\bm{p}}\,,\label{eqn: matrix identity 1}\\
    \sum_{\substack{\bm{p}\in\{0,1\}^n\\p_j=0}}F_{\bm{p}}&=2^{n-1}C_j\,.\label{eqn: matrix identity 2}
\end{align}
Indeed, using the mixed product property of Kronecker products
\begin{equation}
    F_{\bm{p}}H_\ell=\bigotimes_{s=1}^n\left(\begin{bmatrix}1&(-1)^{p_s}\\(-1)^{p_s}&1\end{bmatrix}\cdot
    \begin{bmatrix}1-\delta_{s\ell}&\delta_{s\ell}\\\delta_{s\ell}&1-\delta_{s\ell}\end{bmatrix}\right)
    =\bigotimes_{s=1}^n\begin{cases}\begin{bmatrix}(-1)^{p_\ell}&1\\1&(-1)^{p_\ell}
    \end{bmatrix}
         &(s=\ell)\,,
        \\ \begin{bmatrix}1&(-1)^{p_s}\\(-1)^{p_s}&1\end{bmatrix}  & (s\neq\ell)\,.
    \end{cases}
\end{equation}
By the bilinearity of Kronecker products, we can now extract a factor of $(-1)^{p_\ell}$ to get
\begin{equation}
    F_{\bm{p}}H_\ell=(-1)^{p_\ell}\;\bigotimes_{s=1}^n\;\begin{cases}\begin{bmatrix}1&(-1)^{p_\ell}\\(-1)^{p_\ell}&1
    \end{bmatrix}
        & (s=\ell)
        \\ \begin{bmatrix}1&(-1)^{p_s}\\(-1)^{p_s}&1\end{bmatrix} & (s\neq\ell)
    \end{cases} \quad=(-1)^{p_\ell} F_{\bm{p}}\,.
\end{equation}
Now, fix some $k\in\{1,\ldots,n\}$. Then, we can split the sum
\begin{equation}\label{eqn: Fp sum derivation}
    \begin{split}
        \sum_{\substack{\bm{p}\in\{0,1\}^n\\p_j=0}}F_{\bm{p}}&=\sum_{\substack{\bm{p}\in\{0,1\}^n\\p_j=0}}\bigotimes_{\ell=1}^n\begin{bmatrix}1&(-1)^{p_\ell}\\(-1)^{p_\ell}&1\end{bmatrix}\\
        &=\sum_{\substack{\bm{p}\in\{0,1\}^n\\p_j=p_k=0}}\bigotimes_{\ell=1}^n\begin{bmatrix}1&(-1)^{p_\ell}\\(-1)^{p_\ell}&1\end{bmatrix}+\sum_{\substack{\bm{p}\in\{0,1\}^n\\p_j=0,p_k=1}}\bigotimes_{\ell=1}^n\begin{bmatrix}1&(-1)^{p_\ell}\\(-1)^{p_\ell}&1\end{bmatrix}.
    \end{split}
\end{equation}
For a given choice of $p_1\ldots,\hat{p}_k,\ldots,p_n\in\{0,1\}$, the only difference between the term from the sum on the left and the term from the sum on the right is in the $k$th $2\times2$ matrix appearing in the Kronecker product. Consequently, we can use the bilinearity of Kronecker products to write
\begin{equation}
\begin{split}
    &\bigotimes_{\ell=1}^{k-1}\begin{bmatrix}1&(-1)^{p_\ell}\\(-1)^{p_\ell}&1\end{bmatrix}\otimes\begin{bmatrix}
        1&1\\1&1\end{bmatrix}\otimes\bigotimes_{\ell=k+1}^n\begin{bmatrix}1&(-1)^{p_\ell}\\(-1)^{p_\ell}&1\end{bmatrix}\\
        & \quad +\bigotimes_{\ell=1}^{k-1}\begin{bmatrix}1&(-1)^{p_\ell}\\(-1)^{p_\ell}&1\end{bmatrix}\otimes\begin{bmatrix}
        1&-1\\-1&1\end{bmatrix}\otimes\bigotimes_{\ell=k+1}^n\begin{bmatrix}1&(-1)^{p_\ell}\\(-1)^{p_\ell}&1\end{bmatrix}\\
        =&2\bigotimes_{\ell=1}^{k-1}\begin{bmatrix}1&(-1)^{p_\ell}\\(-1)^{p_\ell}&1\end{bmatrix}\otimes\begin{bmatrix}
        1&0\\0&1\end{bmatrix}\otimes\bigotimes_{\ell=k+1}^n\begin{bmatrix}1&(-1)^{p_\ell}\\(-1)^{p_\ell}&1\end{bmatrix},
\end{split}
\end{equation}
which allows us to re-write \reef{eqn: Fp sum derivation} in the form
\begin{equation}
     \sum_{\substack{\bm{p}\in\{0,1\}^n\\p_j=0}}F_{\bm{p}}=2\sum_{\substack{p_1,\ldots,\hat{p}_k,\ldots,p_n\in\{0,1\}\\p_j=0}} \;\bigotimes_{\ell=1}^n \;
     \begin{cases}\mathbbm{I}_{2\times2}, \ &(\ell=k)\\\begin{bmatrix}1&(-1)^{p_\ell}\\(-1)^{p_\ell}&1\end{bmatrix}\;\; &(\ell\neq k)\end{cases}.
\end{equation}
Repeating this process for each $k\in\{1,\ldots,n\}\backslash\{j\}$, we find that
\begin{equation}
    \sum_{\substack{\bm{p}\in\{0,1\}^n\\p_j=0}}F_{\bm{p}}=2^{n-1}\bigotimes_{\ell=1}^n\begin{cases}\mathbbm{I}_{2\times2}, & (\ell\neq j)\\
    \begin{bmatrix}1&1\\1&1\end{bmatrix} & (\ell=j)\end{cases}
    \quad=2^{n-1}C_j\,.
\end{equation}
So, we have proven the identities \reef{eqn: matrix identity 1} and \reef{eqn: matrix identity 2}. Using identity \reef{eqn: matrix identity 1}, we see that
\begin{equation}
\begin{split}
    N_jA^{(X)}_{\alpha+1}&=\frac{1}{2^{n-1}}\sum_{\substack{\bm{p}\in\{0,1\}^n\\p_j=0}}\frac{1}{1+\alpha+\sum_{\ell=1}^n(-1)^{p_\ell}\xi_\ell}F_{\mathbf{p}}\cdot\bigg[(1+\alpha)\mathbbm{I}_{2^n}+\sum_{\ell=1}^n\xi_\ell H_\ell\bigg]\\
    &=\frac{1}{2^{n-1}}\sum_{\substack{\bm{p}\in\{0,1\}^n\\p_j=0}}\frac{1}{1+\alpha+\sum_{\ell=1}^n(-1)^{p_\ell}\xi_\ell}\bigg[(1+\alpha)F_{\mathbf{p}}+\sum_{\ell=1}^n(-1)^{p_
    \ell}\xi_\ell F_{\mathbf{p}}\bigg]\\
    &=\frac{1}{2^{n-1}}\sum_{\substack{\bm{p}\in\{0,1\}^n\\p_j=0}}F_{\mathbf{p}}\,,
\end{split}
\end{equation}
which through \eqref{eqn: matrix identity 2} directly gives us
\begin{equation}
    N_jA^{(X)}_{\alpha+1}=C_j\,,
\end{equation}
and so indeed $C_j[A^{(X)}_{\alpha+1}]^{-1}=N_j$ (assuming invertibility of $A^{(X)}_{\alpha+1}$). We have thus found the following expression for $M^{\text{ext}}_j$,
\begin{equation}
\begin{split}
    M^{\mathrm{ext}}_j&=iB_j-i\frac{\nu_j}{k_j}C_j[A^{(X)}_{\alpha+1}]^{-1}\\&=i B_j-\frac{i\nu_j}{2^{n-1}k_j}\sum_{\substack{\bm{p}\in\{0,1\}^n \\ p_j=0}}\frac{1}{1+\alpha+\sum_{\ell=1}^n(-1)^{p_\ell}\xi_\ell}F_{\bm{p}}\cdot G\,.
\end{split} 
\end{equation}
In the case $n=j=1$, this formula reproduces the result on the left of \eqref{eq:M def}.

%%%%%%%%%%%%%%%%%%%%%%%%%%%%%%%%%%%%%%%%%%%%%%%%%%%%%%%%%%%%%%%

\section{Repeated Shifts}\label{app:Repeated Shift}
Through repeated application of weight-shifting matrices, we can explore the behaviour at large deviations from conformally coupled weights. This complements the expansion presented in~\cite{massivekinflow} for small $\xi$ as it permits the extension of such an expansion to centre around larger weights. Starting from a conformally coupled contact diagram and then repeatedly shifting the same line $N$ times we find that the master integrals take the form
\ba
    \vec I_{\alpha,\frac{1}{2}+N}=\frac{i^{-N}N!}{2(\alpha+1)_{N}}\begin{bmatrix}
        P_{N}^{(-\alpha,\alpha)}\left(\frac{X}{k}\right)-\frac{N+\alpha}{N}P_{N-1}^{(-\alpha,\alpha)}\left(\frac{X}{k}\right)\\
        P_{N}^{(-\alpha,\alpha)}\left(\frac{X}{k}\right)+\frac{N+\alpha}{N}P_{N-1}^{(-\alpha,\alpha)}\left(\frac{X}{k}\right)
    \end{bmatrix}\psi_{\alpha,\frac{1}{2}}\,,
\ea
where the $P^{(\alpha,\beta)}_N(x)$ are Jacobi polynomials and $(\,\cdot\,)_N$ is the Pochhammer symbol. Adding these entries we compute this shifted contact diagram to be
\ba
    \psi_{\alpha,\frac{1}{2}+N}&=\frac{i^{-N}N!}{(\alpha+1)_N}P_N^{(-\alpha,\alpha)}\left(\frac{X}{k}\right)\psi_{\alpha,\frac{1}{2}}\\&=\frac{C_{\alpha,\frac{1}{2}}(1-\alpha)_N}{i^N(\alpha+1)_N} \,  {}_2F_1\bigg[\begin{array}{c}
1-\alpha+N\ ,\  -N-\alpha \\[-1pt]
1-\alpha
\end{array}\bigg\rvert \, \frac{k-X}{2k}\,\bigg]\,,
\ea
where the function $C_{\alpha,\nu}$ was originally defined in \eqref{eq:Contact Coeff}. Analytically continuing this wavefunction coefficient beyond integer $N$ to the complex numbers we find
\ba
\psi_{\alpha,\frac{1}{2}+N}=e^{-i\pi N}\frac{\sin(\pi(\alpha+N))}{\sin(\pi\alpha)}C_{\alpha,\nu}\left(1-\frac{\xi}{\alpha}\right)\,  {}_2F_1\bigg[\begin{array}{c}
1-\alpha+N\ ,\  -N-\alpha \\[-1pt]
1-\alpha
\end{array}\bigg\rvert \, \frac{k-X}{2k}\,\bigg]\,.
\ea
This is equal to the exact result~\eqref{eq:Explicit Contact} up to the complex prefactor which is equal to $1$ for integer $N$. We thus see that this weight-shifting approach is incapable of fully fixing $\psi_{\alpha,\frac{1}{2}+N}$ for non-integer weights given just the conformally coupled data. However, it can extract the functional dependence in terms of the kinematic variables.

\vskip4pt

We can similarly repeatedly shift the bulk-to-bulk propagator in an all-conformally coupled single exchange diagram. In this case we find
\ba
    \psi_{\alpha_1,\alpha_2,\frac{1}{2}+N}=&\frac{(N!)^2}{(\alpha_1+1)_N(\alpha_2+1)_N}\left[P^{(-\alpha_1,\alpha_1)}_N\left(\frac{X_1}{Y}\right)P^{(-\alpha_2,\alpha_2)}_N\left(-\frac{X_2}{Y}\right)\psi_{\alpha_1,\alpha_2,\frac{1}{2}}^{+-}\right.\\+&\left.P^{(-\alpha_1,\alpha_1)}_N\left(-\frac{X_1}{Y}\right)P^{(-\alpha_2,\alpha_2)}_N\left(\frac{X_2}{Y}\right)\psi_{\alpha_1,\alpha_2,\frac{1}{2}}^{-+}\right]+F_NJ_{\alpha_1,\alpha_2,\frac{1}{2}}\,.
\ea
However, we were unable to find a closed form expression for the set of functions $F_N$. The first two such functions are
\ba
    F_1&=\frac{4(X_1+X_2)}{Y(1+\alpha_1)(1+\alpha_2)}\,,\\
    F_2&=-\frac{12(X_1+X_2)(3X_1X_2-Y^2(1+\alpha_1)(1+\alpha_2))}{Y^3(1+\alpha_1)(2+\alpha_1)(1+\alpha_2)(2+\alpha_2)}\,.
\ea
If one could find a closed form expression for $F_N$ and analytically continue it, this could provide an alternative approach to computing this arbitrary mass exchange diagram, up to some overall scaling that is independent of the kinematics.

%%%%%
\section{\boldmath Cosmological Collider with $\mathcal{M}$}\label{app:Massless Alt}
In this appendix, we present expressions for the diagram in~\eqref{eqn: massless arbitrary mass exchange} in terms of the shifted master integrals $\psi_{\alpha+2,\alpha+2,\mathbf{\frac{1}{2}},\nu}^{\pm\pm}$ and the collapsed term $J_{\alpha+2,\alpha+2,\mathbf{\frac{1}{2}}}$,
\ba\label{eqn: arbitrary exchange massless legs 1}
    \psi_{\alpha,\alpha,\mathbf{\frac{3}{2}},\nu}&=\frac{1}{e_4\left(X_1^{(+)}X_1^{(-)}X_2^{(+)}X_2^{(-)}\right)^2}\left(\sum_{a,b=\pm}\tilde{\mathcal{F}}_1^a\tilde{\mathcal{F}}_2^b\psi_{\alpha+2,\alpha+2,\mathbf{\frac{1}{2}},\nu}^{ab}+\frac{4Y\tilde{\mathcal{G}}}{k_T^3}J_{\alpha+2,\alpha+2,\mathbf{\frac{1}{2}}}\right),
\ea
where we have defined the functions
\ba
    \tilde{\mathcal{F}}_i^\pm&={X_i^{(-)}}^2\left(k_{2i-1}k_{2i}(1+\alpha)(2+\alpha)-X_i^{(+)}(\pm X_i(1+\alpha)-Y)\right)\\&-X_i^{(+)}\left(\pm X_iX_i^{(+)}X_1^{(-)}-k_{2i-1}k_{2i}(\pm X_i(3+2\alpha)-Y)\right)\xi-k_{2i-1}k_{2i}X_i^{(+)}X_i^{(-)}\xi^2\,,\\[1.5ex] 
    \tilde{\mathcal{G}}&=\tilde{\mathcal{G}}_{0}(X_1,X_2,Y)+k_1 k_2 \tilde{\mathcal{G}}_{1}(X_1,X_2,Y)+k_3 k_4 \tilde{\mathcal{G}}_{1}(X_2,X_1,Y)+e_4 \tilde{\mathcal{G}}_{2}(X_1,X_2,Y)\,,
\ea
with
\ba
    \tilde{\mathcal{G}}_0&=k_TX_1^{(-)}X_2^{(-)}X_1^{(+)}X_2^{(+)}(k_T^2(Y^2+X_1X_2(1+\alpha+\xi))-2X_1X_2(X_1X_2+Y^2)(2+\alpha))\,,\\[1.5ex] 
    \tilde{\mathcal{G}}_1&=-2X_2Y^4(2+\alpha)+2X_2^3Y^2(3+2\alpha)+2X_1^4X_2(1+\alpha)(1+\alpha+\xi)\\&+X_1(2Y^2(1+\alpha)-X_2^2(2+\alpha+\xi))(2Y^2(2+\alpha)-X_2^2(1+\alpha+\xi))\\&-X_1X_2^2Y^2(\alpha-\xi)(5+3\alpha-\xi)\\&-2X_1^2X_2(Y^2(-1+\alpha+\alpha^2+\xi+\xi^2)+X_2^2(3+2\alpha-3\alpha\xi-4\xi-\xi^2))\\&+X_1^3(X_2^2(1+\alpha+\xi)(6+4\alpha+\xi)-Y^2(6+9\alpha+3\alpha^2+\xi+\xi^2))\,,
    \\[1.5ex]  
    \tilde{\mathcal{G}}_2&=2(-2(X_1X_2+Y^2)^3(1+\alpha)(2+\alpha)(3+2\alpha)\\&+k_T^4(X_1 X_2(1+\alpha)(1+\alpha+\xi)(2+\alpha+\xi)-Y^2(3+2\alpha)(2+3\alpha+\alpha^2+\xi+\xi^2))\\&+k_T^2(X_1X_2+Y^2)(2+\alpha)(X_1X_2(-\alpha-\alpha^2+\xi+\xi^2)+Y^2(10+17\alpha+7\alpha^2+\xi+\xi^2))\,.
\ea
These expressions are significantly more complicated than~\eqref{eqn: arbitrary exchange massless legs 2}. This appears to be a generic feature of this weight-shifting procedure. The final expressions relating master integrals at different weights are more complicated when shifting between different values of $\alpha$. It is possible that there exists a more direct formula, similar to that derived in Appendix~\ref{app: derivation of contact weight-shifting} that manifests the simplicity of these final expressions. Without such a formula for arbitrary diagrams, it is often easier to use $\mathcal{M}$ to perform the shifts as it is computationally much easier to avoid performing a matrix inversion.

\phantomsection
\addcontentsline{toc}{section}{References}
\bibliographystyle{utphys}
{\linespread{1.075}
	\bibliography{Ref.bib}
}
\end{document}